\documentclass[%reprint,
superscriptaddress,twocolumn,
 amsmath,amssymb,
 aps,
prb,
]{revtex4-2}

\usepackage{dsfont}
\usepackage{graphicx}% Include figure files
\usepackage{dcolumn}% Align table columns on decimal point
\usepackage{bm}% bold math
\usepackage{hyperref}% add hypertext capabilities
\usepackage{graphicx}% Include figure files
\usepackage{hyperref}% add hypertext capabilities
\hypersetup{
colorlinks=true,
	linkcolor=nrppurple,
	citecolor=nrppurple,
	urlcolor=nrppurple,
}

\usepackage{subfigure}

\def \Fst{$1^\textrm{st}$ }
\def \Snd{$2^\textrm{nd}$ }
\def \hS{\hat{\mathcal{S}}  }

\def \A{\mathcal{A}}

\newcommand{\ds}{\displaystyle}
\newcommand{\ket}[1]{\left| #1 \right>} % for Dirac bras
\newcommand{\bra}[1]{\left< #1 \right|} % for Dirac kets
\newcommand{\eqcolon}{\ensuremath{\mathrel{=\!\!\mathop{:}}}}

\def \xim{\xi_{\mu}}
\def \xin{\xi_{\nu}}

\def \uS{\underline{\mathcal{S}}}
\def \Tr{\,\textrm{Tr}\,}	
\def\bs#1{\boldsymbol{#1}}
\def\rm#1{\text{#1}}

\def \Trd{$3^\textrm{rd}$ }

\begin{document}
\definecolor{nrppurple}{RGB}{128,0,128}

% \title{A semiclassical treatment of spinor topological effects in adiabatically driven, weakly inhomogeneous insulators under perturbative electromagnetic fields}
\title{A semiclassical treatment of spinor topological effects in driven, inhomogeneous insulators under external electromagnetic fields}

\author{Ioannis Petrides}
\affiliation{Harvard John A. Paulson School of Engineering and Applied Sciences, Harvard University, MA 02138, Cambridge, USA}
\author{Oded Zilberberg}
\affiliation{Department of Physics, University of Konstanz, D-78457 Konstanz, Germany}

%author order TBD -- we can come back to this..


\begin{abstract}
    Introducing internal degrees of freedom in the description of topological insulators has led to a myriad of theoretical and experimental advances.
    Of particular interest are the effects of periodic perturbations, either in time or space, as they considerably enrich the variety of electronic responses, with examples such as Thouless's charge pump and its higher dimensional cousins, or, higher-order topological insulators.
    Here, we develop a semiclassical approach to transport and accumulation of general spinor degrees of freedom, such as physical spin, valley, or atomic orbits, in adiabatically driven, weakly inhomogeneous insulators of dimensions one, two and three under external electromagnetic fields.
    Specifically, we focus on physical spins and derive the spin current and density up to third order in the spatio-temporal modulations of the system.
    We, then, relate these contributions to geometrical and topological objects -- the spin-Chern fluxes and numbers -- defined over the higher-dimensional phase-space of the system, i.e., its combined momentum-position-time coordinates.
	Furthermore, we provide a connection between our semiclassical analysis and the modern theory of multipole moments by introducing spin analogues of the electric dipole, quadrupole and octapole moments.
	The results are showcased in concrete tight-binding models where the induced responses are calculated analytically.
	
\end{abstract}
\maketitle
%\tableofcontents
%\tableofcontents

\section{Introduction}\label{sec:Intro}

The topological and geometrical aspects of condensed matter systems have yet to be fully explored, even at the single-particle level~\cite{hasan2010rev, qi2011topological, ozawa2019topological}. 
%Homogeneous insulators provide the simplest platform for probing such aspects with electromagnetic fields, since topology is only defined in momentum space (i.e., the material is homogeneous in space) and non-topological contributions from the Fermi surface vanish (i.e., Fermi level in mobility gap).
A prominent platform for probing such physics with electromagnetic fields involves spatially-homogeneous electronic insulators, due to the fact that contributions from the Fermi surface vanish.
In such gapped systems, the electronic spectrum can be associated with a global quantity defined over the entire momentum space -- the topological index. 
Interestingly, this index manifests in quantized responses to the applied electromagnetic fields, e.g., the Hall effect~\cite{Klitzing:1980PRL,thouless1982quantized} or the Streda response~\cite{streda1982theory}.
%Bulk and boundary properties that dependent on the topological index remain robust against disorder that preserves the topology.
Following several decades of research, many topological aspects of momentum space are well understood using rigorous mathematical methods, such as K-theory~\cite{bellissard1992gap, bellissard2001gap,shiozaki2017topological,kitaev2009periodic}, non-linear sigma model analysis~\cite{chiu2016classification, ryu2010topological, altland1997nonstandard}, and dimensional reduction~\cite{QiZhang,qi2011topological,teo2010topological}, that rely on a combination of local symmetries~\cite{ryu2010topological, altland1997nonstandard,bellissard1992gap,kitaev2009periodic}, symmorphic or nonsymmorphic crystalline symmetries~\cite{shiozaki2017topological,chiu2016classification,fu2011topological,alexandradinata2016topological}, or even quasiperiodicity~\cite{bellissard2000hull,bellissard2001gap,Kraus2012a,kraus2016quasiperiodicity,kruthoff2017topological} to classify electronic lattices. % .

In recent years, theoretical and experimental studies in ultracold atoms~\cite{Lohse2018,Lohse2016,Nakajima2016,Goldman:2016NatPhys, cooper2018,mei2019topological}, photonics~\cite{Kraus2012a,Verbin2015,lu2016topological, khanikaev2017two, ozawa2019topological,benalcazar2020higher,Zilberberg2018}, mechanical systems~\cite{serra2018observation,apigo2018topological,cheng2020demonstration,long2019floquet,xia2020experimental,grinberg2020robust,rosa2019edge,tsai2019topological}, electrical circuits~\cite{liu2019topologically,peterson2018quantized,imhof2018topolectrical,serra2019observation,ezawa2019electric,yu20194d}, and Moire heterostructures~\cite{su2020topological} have shown enormous capabilities in simulating exotic quantum phenomena.
In particular, the induced responses that arise in systems subject to time-dependent modulations were shown to depend on topological aspects that go beyond the traditional momentum-space description.
%An archetypical example is Thouless's 1D pump~\cite{thouless1982quantized, Thouless83,kraus2012ye,Verbin2015,Lohse2016,Nakajima2016}, where the adiabatic and periodic evolution of the system's parameters results in a quantized charge transport that is related to a \Fst Chern number defined over the combined momentum-time manifold. 
An archetypical example is Thouless's 1D charge pump~\cite{thouless1982quantized, Thouless83,kraus2012ye,Kraus2012a,Verbin2015,Lohse2016,Nakajima2016}, where the adiabatic and periodic modulation of the system's parameters results in the transport of a quantized amount of charge across the otherwise insulating bulk; such quantization was shown to be related to a \Fst Chern number defined over the combined momentum-time manifold.
Its extension to two and three dimensions led to topological charge pumps with a \Snd and \Trd Chern number response, as well as to a plethora of associated boundary physics~\cite{qi2011topological, kraus2013,Lohse2018,Zilberberg2018,petrides2018six}.

Complimentary to topological charge pumps are the newly found higher-order TIs, where the ground state is characterized by the existence of fractional boundary charges with co-dimensions $>1$~~\cite{ryu2009masses,lin2017topological,hashimoto2017edge,langbehn2017reflection,benalcazar2017quantized,trifunovic2018higher,geier2018second,cualuguaru2019higher,zhao2020electric,song2017d}.
Such states can be classified by the electric multipole moments that take quantized values when constrained by symmetries. 
The appearance of nontrivial electric multipoles and localized charges finds numerous manifestations in crystaline materials~\cite{schindler2018higher,wang2018higher,ezawa2018minimal}, as well as in photonic lattices~\cite{Zilberberg2018,cerjan2020observation}, metamaterials~\cite{serra2018observation,kovsata2021second}, electrical circuits~\cite{imhof2018topolectrical}, and, superconductors~\cite{ghorashi2019second,teo2010topological}.

% both in studying foundamental physics with dark matter detectors~\cite{marsh2019proposal,ringwald2012exploring}, 
% as well as engineering next-generation devices of topological quantum states~\cite{hasan2010rev,qi2011topological}.

% in Chemical engineering~\cite{bradlyn2017topological}, or in realizing next-generation devices of topological quantum states~\cite{hasan2010rev,qi2011topological}.%,lutchyn2010majorana,oreg2010helical

An alternative description of higher-order TIs is found within the semiclassical theory, where physical observables were shown to depend on the topological aspects of the entire phase-space, i.e., the combined momentum-position-time manifold~\cite{Niu1995, Sundaram1999, Xiao2010, Xiao2009, Gao2014, Price2015, price2016measurement,petrides2018six,Lee18}.
For example, charge transport and accumulation were shown to depend on geometrical quantities, called Chern fluxes, that become quantized and fractional when global symmetries are imposed; these are related to quantized changes of the electric multipoles and, hence, to localized charges~\cite{petrides2020higher}.

% The accumulation of charge at the boundaries of higher-order TIs is related to geometrical quantities defined over the momentum-position manifold, the Chern fluxes, that become quantized and fractional when global symmetries are imposed~\cite{petrides2020higher}.
% The emergence of nontrivial electronic responses associated to the charge degree of freedom are well understood within the semiclassical theory where physical observables depend on the topological aspects of the entire phase space, i.e., the combined momentum-position-time manifold~\cite{Niu1995, Sundaram1999, Xiao2010, Xiao2009, Gao2014, Price2015, price2016measurement,petrides2018six,Lee18}.

Generalizing the description to other internal degrees of freedom beyond charge offers new possibilities in engineering next-generation devices using topological quantum states.
For example, the coherent control and manipulation of physical spins finds numerous applications in spintronics and has motivated the search of dissipationless spin-currents in quantum dot structures~\cite{watson2003experimental}, in spin-Hall systems, such as doped GaAs~\cite{murakami2003dissipationless,sinova2004universal,kato2004observation,wunderlich2005experimental}, and in 1D topological spin pumps~\cite{sharma2001quantum,mucciolo2002adiabatic,zhou2004topological,shindou2005quantum,fu2006time}.
The latter is the analogue of Thouless's 1D charge pump with a direct relation between spin transport and a \Fst spin-Chern number defined over the system's parameter space.
However, largely unexplored are the topological signatures of spin observables combined with spatio-temporal modulation in higher-dimensions. %and the relation of spin observables to the topological aspects of phase-space.
% Extending the study to include spin degrees of freedom is expected to open new possibilities in engineering topological responses. %associated to general spin degrees of freedom; as foretold by their analogue systems with nontrivial charge responses discussed above.

Here, we derive the transport and accumulation of general degrees of freedom, dubbed ``spinor" degrees, in adiabatically driven, weakly inhomogeneous insulators in dimensions one, two and three, under external electromagnetic fields.
We start by reviewing the semiclassical description of crystalline materials and show how the equations of motion of an electron wavepacket lead to charge transport and accumulation.
We then extend this framework to include spinor degrees of freedom and derive the spinor-charge transport and accumulation up to third order in perturbation theory.
We find that these are related to topological and geometrical quantities -- the spinor-Chern numbers and fluxes -- that are defined over the entire phase-space of the system.
Similar to charge responses~\cite{petrides2018six}, we obtain generalizations of topological spinor pumps, the spinor-Hall effect, spinor-higher-order TIs, and spinor-Axion responses.
Finally, we decompose the derived corrections into spinor analogues of the electric multipole moments, thus, establishing a direct relation between macroscopic properties of the material and the spinor-topological aspects of phase-space.
We showcase our results in concrete tight-binding Hamiltonians, where we focus on physical spin.

\section{Semiclassical approach}\label{sec:Theory}
In this Section, we review the semiclassical description of electrons in insulating materials under general perturbing fields and show how the geometrical properties of phase-space manifest as corrections to charge transport and accumulation~\cite{Niu1995, Sundaram1999, Xiao2010, Xiao2009, Gao2014, Price2015, price2016measurement,petrides2018six,Lee18}.
Depending on the dimensionality of the system, the equations of motion will include effects up to third order in the perturbing fields~\cite{petrides2018six,thouless1982quantized,Price2015,prodan2016bulk}, as it is only at this order that electronic responses of 3D materials can be completely characterized. 
Once the semiclassical theory of the charge degree of freedom is reviewed, we will extend it to general spinor degrees and straightforwardly apply it to dimensions one, two and three in Section~\ref{sec:Applications}.

The semiclassical theory offers an intuitive picture for describing transport and accumulation of charged particles moving in insulators that are subject to weak spatio-temporal modulations.
It describes the particles as wavepackets that adiabatically move in phase-space with respect to a local eigenbasis, see Fig.~\ref{fig:figsemi}(a).
Specifically, the wavepacket is assumed to have a well-defined center-of-mass coordinates $\bm \xi = (\bs r, \bs k)$, where $\bs r$ denotes the position and $\bs k$ the crystal's quasimomentum, such that its dynamics can be perturbatively expanded at small distances as
\begin{eqnarray}
    \hat{H} \approx  \hat{H}_0 +\hat{H}'\,.\label{eq:HamSemi}
\end{eqnarray}
Here, $\hat{H}_0$ is the unperturbed Hamiltonian and $\hat{H}'$ are higher-order corrections.
In this case, the wavepacket is built directly from the $N$ eigenstates
\begin{eqnarray}
    \ket{{n}(\bs{\xi})} \approx \ket{n_0} +\ket{n'}\notag
\end{eqnarray} 
of a set of isolated energy bands of $\hat{H}$, where $\ket{n_0}$ are the eigenstates of $\hat{H}_0 $ and $\ket{n'}$ are higher-order corrections.

For more details on the construction of the wavepacket and the derivation of its equations of motion up to third-order, see Refs.~\cite{Niu1995,Chang1995,culcer2005coherent,  Sundaram1999, Xiao2010, Xiao2009, Gao2014, Price2015, price2016measurement,petrides2018six,Lee18} and Appendix~\ref{apx:semi}.
Here, we directly use the resulting velocity and force equations describing the center-of-mass evolution
\begin{eqnarray}
	\begin{array}{rl}
		\ds\underline{\dot{r}} ^i &\ds= \partial_ {k_i}  { \mathcal{E}}  - {\Omega}_{k_i k_j}\dot{\underline{k}} _j- {\Omega}_{k_i r_j} \dot{\underline{r}}_j -{\Omega}_{k_i t}\,,  \\\\
		\ds\underline{\dot{k}} ^{i} &\ds= \partial_ {r_i}  { \mathcal{E}}  - {\Omega}_{r_i k_j}\dot{\underline{k}}_j- {\Omega}_{r_i r_j}\dot{\underline{r}}_j -{\Omega}_{r_i t}\,,
		\label{eq:EoM nonlinear}
	\end{array}	
\end{eqnarray}
where $\underline{{\bm r}}_{nm} =  \bra{{n} } \hat{\bm r} \ket{{m}}$ and $\underline{{\bm k}}_{nm} = \bra{{n} } \hat{\bm k} \ket{{m}}$ are the matrix representations of the wavepacket's center-of-mass position and momentum, the energy dispersion $ \mathcal{{{E}}}^{nm} = 	\bra{{n} (\bs{\xi})}{\hat{H}_0 + \hat{H}'}	\ket{{m} (\bs{\xi})}$ is calculated up to a sufficiently high order in perturbation theory, and Einstein's summation convention is assumed.
The curvature tensors are defined as 
\begin{eqnarray}
	{\Omega}_{\xim \xin} &=&  \partial_{\xim}{ \mathcal{A}}_{\xin }  -  \partial_{\xin}{ \mathcal{A}}_{\xim} \,, \label{eq:berry2form}
\end{eqnarray}
where
\begin{eqnarray}
	{ \mathcal{A}}^{nm} _{\xi_\mu} = \bra{{n} (\bs{\xi})}{i \partial_{\xi_\mu}}\ket{{m} (\bs{\xi})}\,,
	\label{eq:conn}
\end{eqnarray}
is the so-called connection and $\xi_\mu$ denotes the $\mu$-th coordinate in phase-space.
Depending on the particular directions involved, we dub $ \Omega_{k_i k_j}$ the ``momentum (Berry) curvature", $ \Omega_{r_i r_j}\equiv B_{ij}$ the ``position (magnetic) curvature", while $ \Omega_{k_i r_j}$, $ \Omega_{k_i t}$, and $ \Omega_{r_i t}$ are dubbed the mixed momentum-position, momentum-time, and position-time curvatures.

The equations of motion~\eqref{eq:EoM nonlinear} exhibit the usual dependence on the group velocity $\frac{\partial  \mathcal{E}}{ \partial {k_i} }$~\cite{Karplus:1954PR} and force $\frac{\partial  \mathcal{E}}{ \partial {r_i} }$, while the curvature tensors ${\Omega}_{\xi_\mu \xi_\nu} $ appear as ``anomalous velocity" and ``anomalous force" terms that modify the trajectories of the wavepacket depending on the geometrical structure of phase-space~\cite{PricePRA2012,Cominotti2013,Wimmer:2017NatPhys}. 
For example, the anomalous velocity ${\Omega}_{k_i k_j}\dot{k} _j$ can be understood as a momentum-space analogue of the magnetic Lorentz force. %, in which the Berry curvature acts like a magnetic field in momentum space
This gives rise to the quantum Hall effect~\cite{Klitzing:1980PRL,thouless1982quantized} and can be used to map out the distribution of the Berry curvature over energy bands~\cite{PricePRA2012,Cominotti2013,Wimmer:2017NatPhys}.
%Furthermore, depending on the dimensionality, these anomalous terms lead to topological charge pumps with their associated \Fst, \Snd and \Trd Chern number responses, in addition to the standard quantum Hall effect in two dimensions and the magneto-electric effect in time-modulated 3D systems~\cite{qi2011topological}.

Assuming that energy bands are uniformly filled up to some spectral gap, the associated charge density and current reads
\begin{equation}
	\displaystyle \rho_{\rm{particle}}= \int\limits_{\mathbb{T}^d} \frac{\mathrm{d}^d {k}}{\left(2\pi\right)^d}\Tr D(\bs{\xi}) \,,
	\label{eq:density}
\end{equation} 
\begin{equation}
	\displaystyle \bm j _{\rm{particle}}= \int\limits_{\mathbb{T}^d} \frac{\mathrm{d}^d {k}}{\left(2\pi\right)^d}\Tr D(\bs{\xi}) \underline{\dot{\bm r}}	\,,
	\label{eq:current}
\end{equation} 
respectively, where the integral runs over the entire $d$-dimensional Brillouin zone denoted by the $d$-torus $\mathbb{T}^d$, the trace is performed over the set of occupied states, and $D(\bs{\xi})$ is the modified density of states.
The latter is a consequence of the underlying geometry of phase-space, as it takes into account the change in the number of available states when nontrivial curvature tensors are included~\cite{Xiao2005,Duval2005,Bliokh2006,Gosselin2006, Price2015, price2016measurement}, see Fig.~\ref{fig:figsemi}(b).
This change can be classically understood from Liouville's theorem, which states that if the dynamics are Hamiltonian, the phase-space volume element is conserved when transforming from \textit{canonical} to \textit{physical} coordinates. 
Using generalized Peierls substitutions~\cite{Xiao2005,Duval2005,Bliokh2006,Gosselin2006}, we note that the physical coordinates $\bm \xi =(\bm r , \bm k)$ are related to the canonical coordinates $ \bm \Xi =(\bm R , \bm K)$ by  ${\bs r} = {\bs R} - \bm{ \mathcal{A}}_k$ and ${\bs k} = {\bs K} - {\bs \A}_r$, where ${\bs \A}_r$ (${\bs \A}_k$) is the position (momentum) connection, cf. Eq.~\eqref{eq:conn}.
The extent by which the physical coordinates deviate from being canonical is quantified by the curvature tensors $	{\Omega}_{\xim \xin} $, cf. Eq.~\eqref{eq:berry2form}, and the change of phase-space volume element is described by the Jacobian of the transformation~\cite{Duval2005}, given by
\begin{align}
	D(\bs{\xi}) = \sqrt{\text{det}( \left(\begin{matrix}
		\underline{\Omega}^{(r)} & -\mathds{1}-\underline{\Omega}^{(rk)}\\
		\mathds{1}+\underline{\Omega}^{(rk)}  & \underline{\Omega}^{(k)}
	\end{matrix}\right)}\,,
\label{eq:dos}
\end{align}
where $\mathds{1}$ is the identity matrix and $\underline{\Omega}^{(k)}$, $\underline{\Omega}^{(r)}$, and $\underline{\Omega}^{(rk)}$ are antisymmetric matrices with components ${\Omega}_{k_i k_j}$, ${\Omega}_{r_i r_j}$, and ${\Omega}_{r_i k_j}$ respectively. 
%Note that the curvature tensors ${\Omega}_{\xi_i \xi_j}$ are still matrices in the basis of occupied Bloch states and Eq.~\eqref{eq:dos} is calculated with respect to every such component.
The total change in phase-space volume is found by tracing $D(\bs{r},\bs{k}) $ over the occupied energy bands, as done in Eq.~\eqref{eq:density}.

In fermionic systems, the particle density $ \rho_{\rm{particle}}$ is proportional to the charge accumulation induced by charged carriers. 
However, we emphasize that the semiclassical formalism can straightforwardly be applied to a uniformly filled set of bands of bosons~\cite{price2016measurement}.
As we will see in Section~\ref{sec:Applications}, the semiclassical approximation of the particle density at first-order in perturbing fields gives rise to a quantised particle accumulation with co-dimensions $1$, i.e., the ground state supports states that are localized in one dimension but extended in the other directions.
This is closely related to the soliton solutions found in Ref.~\cite{jackiw1976solitons} in the context of high-energy physics.
At second-order it results in the Streda formula~\cite{streda1982theory}, and in a quantised charge accumulation with co-dimensions $2$~\cite{schindler2018higher,ryu2009masses,petrides2020higher,teo2010topological,Hou07,kosata21}.
Finally, third-order terms give rise to Axion responses in the spatial domain~\cite{QiZhang,qi2011topological,nenno2020axion}, and quantised charge accumulation with co-dimensions $3$~\cite{benalcazar2017quantized}.

Next, the particle current of Eq.~\eqref{eq:current} is calculated by integrating the corresponding velocity $\underline{\dot{r}} ^i$ over the entire $d$-dimensional Brillouin zone, weighted by the density of states $D(\bs{\xi})$.
The velocity of the wavepacket in phase-space is found by recursively solving the differential equations~\eqref{eq:EoM nonlinear} up to a particular order, while the density of states is given in Eq.~\eqref{eq:dos}.
The induced corrections are hence classified into density-type, Lorentz-type, or, mixed Lorentz-density-type responses, depending whether they result from the density of states, the velocity, or a combination of the two~\cite{petrides2018six,Lohse2018,Zilberberg2018}.

As we will see in Section~\ref{sec:Applications}, the corrections to the particle current of an insulating ground state give rise to Thouless's 1D charge pump~\cite{thouless1982quantized, Thouless83,kraus2012ye,Kraus2012a,Verbin2015,Lohse2016,Nakajima2016} and to the quantum Hall effect~\cite{Klitzing:1980PRL,thouless1982quantized}; both having a characteristic \Fst Chern number response.
At higher orders, we recover 2D topological charge pumps~\cite{qi2011topological, kraus2013,Lohse2018,Zilberberg2018,benalcazar2020higher} and Axion physics in the temporal domain~\cite{QiZhang,qi2011topological}, where the associated responses are determined by a \Snd Chern number.
Finally, third-order corrections give rise to 3D topological charge pumps and a \Trd Chern number response~\cite{petrides2018six,Lee18}.

%~\cite{qi2011topological, kraus2013,Lohse2018,Zilberberg2018,petrides2018six}~\cite{thouless1982quantized, Thouless83,kraus2012ye,Kraus2012a,Verbin2015,Lohse2016,Nakajima2016}

\subsection{Spinor current and density}

We generalise the semiclassical description of charge transport and accumulation to other quantum numbers, which we generally dub as ``spinor degrees of freedom".
These degrees can represent various particle properties, e.g., charge, physical spin, valley index, or any other internal degree. 
The derivation of the current and density follows a similar procedure, with the difference now that all quantities are defined with respect to the spinor operator $\hat{\mathcal{S}}$
\begin{eqnarray}
	\rho_{\hat{\mathcal{S}}}  =&\displaystyle\int\limits_{\mathbb{T}^d} \frac{\mathrm{d}^d {k}}{\left(2\pi\right)^d} \Tr  \underline{\mathcal{S}} D_{}(\bs{\xi})	\,,\label{eq:DoS general}\\
	\bm j_{\hat{\mathcal{S}}} =&\displaystyle\int\limits_{\mathbb{T}^d} \frac{\mathrm{d}^d {k}}{\left(2\pi\right)^d} \Tr  \mathcal{\underline{S}} D_{}(\bs{\xi}) \underline{\dot{\bm r}}\,,
	\label{eq:current general}
\end{eqnarray}
where $\mathcal{\underline{S}}$ is the matrix representation of the operator $\mathcal{\hat{S}}$ with components $ \mathcal{\underline{S}}^{nm} = e_s \bra{n(\bm \xi)}\hat{\mathcal{S}}\ket{m(\bm \xi)}$, and, $e_s$ is the value of the associated spinor-charge.
% As these expressions depend on the dimensionality of the system, we analyse the resulting terms in increasing order of dimensions.
Even though the formalism is generic to any operator $\hat{\mathcal{S}}$, here, we focus on physical spins in concrete tight-binding models, and analytically calculate the induced responses.
By exploiting the full breadth of these corrections, we engineer dynamical systems, where the quantised spin transport and accumulation are related to nontrivial topological indices defined over the system's phase-space.
	
\subsection{Geometrical definitions}
Before continuing, it is useful to define general geometrical and topological quantities in phase-space that will later appear as physical corrections to spin or charge transport and accumulation.
For a uniformly occupied set of eigenstates, we define the spinor analogues of Chern numbers, sub-Chern numbers, and  Chern fluxes in arbitrary dimensions. %~\cite{thouless1982quantized, Nakahara, chiu2016classification,zhang2001four,Avron1988,QiZhang,prodan2016bulk,Nakahara}

First, the \Fst spinor-Chern number is defined as
\begin{eqnarray}
	c _1 =\!\frac{1}{2 \pi} \int_{\mathbb{T}^2} \mathrm{d}^2 {\xi} \text{Tr}\uS\Omega_{\xim\xin} 
	\,,\label{eq:first_chern} 
\end{eqnarray}
where the integral is taken over a 2D closed surface in the $(\xi_\mu , \xi_\nu)$ plane, denoted here by $\mathbb{T}^2$. 
We note that the degree of freedom $\uS$ is used in a generic way to represent any type of quantum number.
For example, when it corresponds to physical spins, this topological index -- called the \Fst spin-Chern number -- takes integer values and governs the robust quantization of spin-conductance in the 2D spin-Hall effect~\cite{murakami2003dissipationless,sinova2004universal,kato2004observation,wunderlich2005experimental,HughesSpinHall,qi2006topological}, and the quantized spin transport in 1D topological spin pumps~\cite{fu2006time}.
Alternatively, when $\uS$ is the identity, it corresponds to charge, and the above equation is reduced to the well-known \Fst Chern number. 
The latter determines the 2D quantum Hall effect~\cite{Klitzing:1980PRL,thouless1982quantized}, the center-of-mass drift of an atomic cloud~\cite{Aidelsburger:2015NatPhys}, the dynamical vortex trajectories of a quenched cold-atom gas~\cite{tarnowski2017characterizing}, the heating rate of shaken systems~\cite{Tran2017,tran2018quantized,asteria2018measuring}, and the charge transport of 1D topological charge pumps~\cite{Lohse2016, Nakajima2016}. 
%In addition, it has also been proposed to extract the \Fst Chern number from the steady state of driven-dissipative systems~\cite{ozawa2014anomalous,Salerno:2016PRB,Ozawa2016}. 

The \Snd spinor-Chern number emerges in a four-dimensional manifold and it is given by the antisymmetric product of two 2-forms
\begin{align}
	c_2&=\frac{1}{32 \pi^2} \int_{\mathbb{T}^4} \mathrm{d}^4 \xi \epsilon^{\alpha \beta \gamma \delta }  \text{Tr}\uS \Omega_{\xi_\alpha \xi_\beta } \Omega_{\xi_\gamma \xi_\delta} \,, \label{eq:second_chern}
\end{align}
where $\mathbb{T}^4$ denotes the 4D closed manifold and where $ \epsilon^{\alpha \beta \gamma \delta }$ is the Levi-Civita symbol defined in the 4D $\bm \xi$-coordinate space.
When $\uS$ corresponds to the charge degree of freedom, it is exactly the \Snd Chern number appearing in the nonlinear 4D quantum Hall response of a system with four spatial dimensions~\cite{zhang2001four, Price2015,frohlich2000,QiZhang,prodan2016bulk}, in the bulk transport of two-dimensional topological pumps~\cite{Lohse2018}, as well as in the dynamics of internal states in Bose-Einstein condensates~\cite{Sugawa:2016arXiv,Bardyn:2014, Mochol2018}.

Lastly, the relevant topological invariant in a six-dimensional manifold is the \Trd spinor-Chern number
\begin{align}
	c_3 &=   \frac{1}{\left(2\pi\right)^3}\int_{\substack{\mathbb{T}^6}}\mathrm{d}^6 \xi\frac{1}{2^3\cdot 3!}\epsilon^{\alpha\beta\gamma\delta\epsilon\zeta}\text{Tr}\uS\Omega_{\xi_\alpha\xi_\beta} \Omega_{\xi_\gamma\xi_\delta} \Omega_{\xi_\epsilon\xi_\zeta}\,, \label{eq:3rd}
\end{align}
where the 6D $\bs\xi$-coordinate space is denoted by $\mathbb{T}^6$ and where we have introduced the 6D Levi-Civita symbol $\epsilon^{\alpha\beta\gamma\delta\epsilon\zeta}$. 
The \Trd 
spinor-Chern number is inherently a 6D topological invariant as it vanishes for systems with fewer than six dimensions. 
It underlies the 6D quantum Hall effect and it manifests in the charge transport of 3D topological charge pumps~\cite{petrides2018six}.
%Continuing further up in dimensionality, a new Chern number emerges every time the number of dimensions is increased by two, where each successive Chern number can be defined as a higher wedge product of the 2-form curvature~\cite{Nakahara, Chiu:2016RMP,prodan2016bulk}. 

\begin{figure}
	\centering
	\includegraphics[width=1\linewidth]{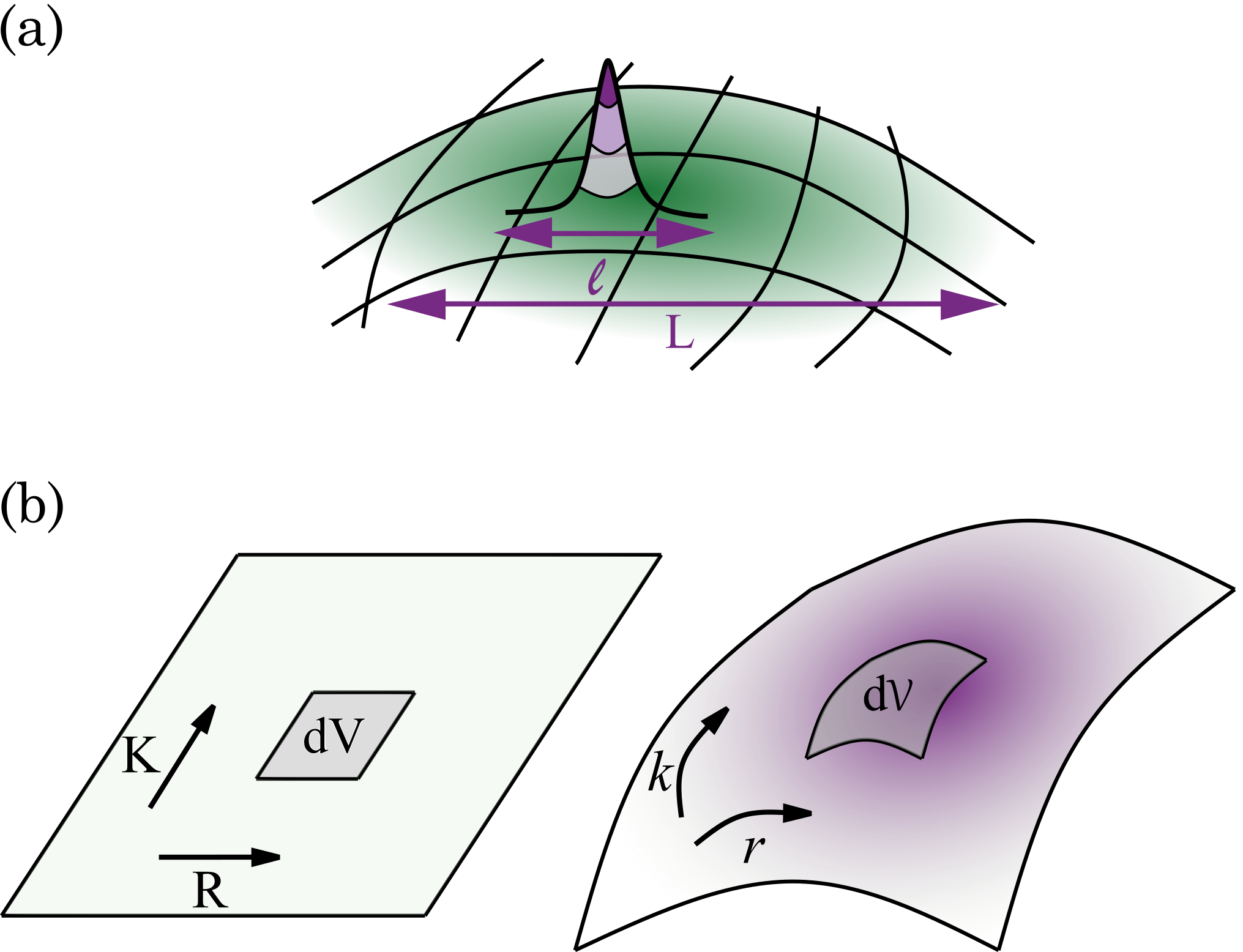}
	\caption{Semiclassical theory. (a) The electron is represented by a wavepacket with well-localized center of mass coordinates, that moves in the phase-space $(\bm r, \bm k)$ of the system. 
	Its trajectory is defined by the anomalous forces and velocities that arise due to the geometry of phase-space, cf. Eq.~\eqref{eq:EoM nonlinear}.
	The length scale $\text{L}$ defined by the perturbing fields is assumed to be much longer than the width $l$ of the wavepacket, such that a local Hamiltonian can always be defined, cf. Eq.~\eqref{eq:HamSemi}.
(b) For canonical coordinates, the phase-space volume element is constant.
When transforming to physical coordinates, it is modified according to the curvature tensors appearing in the equations of motion, cf. Eq.~\eqref{eq:dos}.}
	\label{fig:figsemi}
\end{figure}

For a given set of $\bs\xi$ coordinates, it is important to remember that all lower-dimensional topological indices can still be defined, but now with respect to the various sub-dimensional manifolds~\cite{Nakahara}. 
In practice, each set of states in a $D$-dimensional coordinate space is characterized by a set of \Fst spinor-Chern numbers, associated with each possible 2D plane; a set of \Snd spinor-Chern numbers, associated with each possible 4D subvolume; all the way up to the $D/2$-th spinor-Chern number (where $D$ is even) that characterizes the entire manifold of states.
We dub such lower-dimensional quantities as ``sub-spinor-Chern numbers". 
Importantly, these are not integer-valued as the integrals run over the entire $D$-dimensional space.
Instead, they depend both on the relevant lower-dimensional spinor-Chern numbers as well as on the volume of the coordinate space perpendicular to the selected sub-manifold~\cite{Price2015,price2016measurement}. 

Analogously to the spinor-Chern numbers, the \Fst spinor-Chern flux is defined as
\begin{eqnarray}
	\Phi _1  = \frac{1}{2\pi}\int_{\mathcal{C}} \text{d}\mathcal{C} \Tr \uS \Omega_{\xi_\mu\xi_\nu}\,,
	\label{def:flux1}
\end{eqnarray}
where $\Omega_{\xi_\mu\xi_\nu}$ is the curvature in the $({\xim, \xin})$ plane, and $\mathcal{C}$ is an \textit{open} integration domain with volume element $\text{d}\mathcal{C}$.
This quantity is related to the \Fst spinor-Chern number, but, now the integration domain runs over a sub-volume $\mathcal{C}$ of the entire 2-dimensional manifold.

Going up in dimensionality, the \Snd and \Trd spinor-Chern fluxes are defined as
\begin{eqnarray}
	\ds	\Phi _2 & =&\ds \frac{1}{32\pi^2}\int_{\mathcal{C}} \text{d}\mathcal{C} \epsilon^{\mu\nu \sigma\rho}\Tr \underline{\mathcal{S}}\Omega_{\xi_\mu\xi_\nu}\Omega_{\xi_\sigma\xi_\rho}\,,
		\label{def:flux2}\\
	\ds	\Phi _3 & = &\ds \frac{1}{(2\pi)^3}\int_{\mathcal{C}} \text{d}\mathcal{C}\frac{1}{2^3\cdot 3!} \epsilon^{\mu\nu \sigma\rho\gamma\delta} \Tr \underline{\mathcal{S}}\Omega_{\xi_\mu\xi_\nu}\Omega_{\xi_\sigma\xi_\rho}\Omega_{\xi_\gamma\xi_\delta}\notag\,,\\
		\label{def:flux3}
\end{eqnarray}
where $\mathcal{C}$ is the corresponding integration domain with volume element $\text{d}\mathcal{C}$.
In the definitions of the spinor-Chern fluxes, the integration domain $\mathcal{C}$ does not cover the entire manifold, hence, such expressions are generally not quantized. 
However, as we will later see, global symmetries can constrain the allowed values of these quantities to support only discrete fractions. 

In the following, we will use the term "spin-Chern number" to indicate the case where $\uS$ corresponds to physical spin, the term "Chern number" when $\uS$ corresponds to charge, and "spinor-Chern number" whenever we consider general degrees (analogously for the remaining geometrical definitions). 

\section{Driven, inhomogeneous insulators under electromagnetic fields}\label{sec:Applications}

In this Section, we use the semiclassical approach [cf. Sec.~\ref{sec:Theory}] to calculate the spin transport and accumulation induced by weakly perturbing a crystal in momentum, position, and in time.
In practice, these perturbations correspond to external electromagnetic fields, weak spatial inhomogeneities and adiabatic drives, respectively.
We start by calculating the quantised spin transport in a one dimensional model using our semiclassical approach, where we obtain the same results as originally derived using a quantum mechanical approach in Ref.~\cite{fu2006time}.
Then, we derive the corrections to the spin density and calculate the spin accumulation on a domain wall induced by weakly modulating the Hamiltonian in space.
We showcase the results in a concrete tight-binding model and generalize the description to dimensions two and three. 

To our knowledge, the derivation of spinor, and in particular spin, transport and accumulation in arbitrary dimensions using the semiclassical theory is introduced here for the first time.
At its core, spinor transport and accumulation are proportional to geometrical and topological quantities defined over the system's phase-space -- the spinor-Chern numbers and spinor-Chern fluxes.
We explore the various manifestations of these quantities in driven, inhomogeneous crystals under electromagnetic fields and, in particular, relate them to the spatio-temporal modulations of spinor analogues of the electric multipole moments.
As such, we provide a complete description of noninteracting electrons in perturbed crystalline materials and illuminate a fundamental connection between the topological aspects of phase-space and physical observables.

\subsection{In one dimension}\label{sec:1D}

In the following, we derive the corrections to the spin current and density of a one-dimensional adiabatically driven insulator with weak spatial inhomogeneities using the semiclassical approach of Sec.~\ref{sec:Theory}.
First, we review the calculation of spin transport in a periodically driven Hamiltonian and show its relation to the temporal change of a spin-dipole moment.
Similar to the 1D Thouless charge pump, the spin transport after a cycle is found to be quantised and equal to a \Fst spin-Chern number defined in the phase-space of the system.
We extend these results to derive the spin accumulation on a domain wall created by smoothly modulating the Hamiltonian in position and find that it is related to a geometrical property of phase-space -- the \Fst spin-Chern flux.
Even though such a quantity is generally not quantised, we show that under global symmetry constraints it can become quantised and lead to a fractional spin accumulation localised at the interface.
The relation to the spatial modulation of the spin-dipole moment is also discussed.

\begin{figure}
	\centering
	\includegraphics[width=1\linewidth]{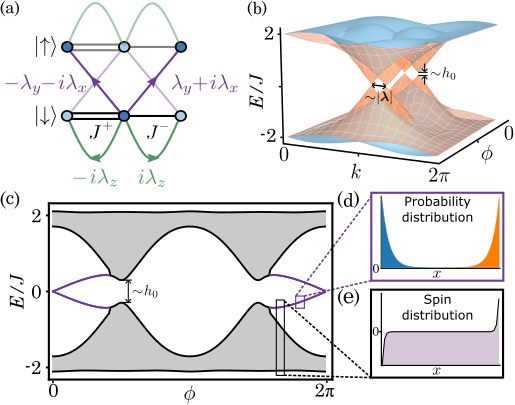}
	\caption{The model Hamiltonian in one dimension. 
		(a) Sketch of the tight-binding model [cf.~Eq.~\eqref{eq:Ham 1D}].
		Electrons with spin $\alpha$ tunnel to the nearest neighbour with amplitude $J^\pm = J \pm \Delta J$, denoted by single/double black lines. 
		The spin-orbit coupling is denoted by the purple and green lines. 
		The light/dark blue sites denote a staggered potential with positive/negative sign.
		(b) The bulk spectrum of $H(k,\phi)$ [cf.~Eq.~\eqref{eq:1DHamMom}] as a function of quasimomentum $\bm k$ and the internal parameter $\phi$.
		The spectrum is composed by four bands that become degenerate at the high symmetry points. 
		(c)  The open boundary spectrum of the Hamiltonian showing bulk (gray), and co-dimension 1 (purple) states.
		(d) The probability distribution of the two lowest energy boundary modes. 
		Each state is occupied at opposite boundaries, resulting in vanishing electric dipole moment.
		(e) The spin density of the ground state at half filling showing spin accumulation at the boundary.
		For the simulations in (c), (d), and, (e) we have used $J/10=\lambda_x =\lambda_y = \lambda_z = J-\Delta J= 0.1$, and $h_0=0.5$, while for (b) we have used $h_0=0.05$.
	}
	\label{fig:fig1d}
\end{figure}

We consider a tight-binding model describing spinful electrons moving on a 1D lattice, see Fig.~\ref{fig:fig1d}(a), with Hamiltonian
\begin{eqnarray}
	H = H_\text{\text{KE}} + H_\text{d} + H_\text{h} +  H_\text{SO}  \,.
	\label{eq:Ham 1D}
\end{eqnarray}
The first term is the kinetic energy, given by 
\begin{eqnarray}
	H_\text{\text{KE}} = J \sum_{n, \alpha}(c^\dagger _{{n} +	1,\alpha} c_{n,\alpha} + \text{h.c})\,,
\end{eqnarray}
where $J$ is the hopping amplitude, $n$ is the position vector, $\alpha=\{\uparrow,\downarrow\}$ the spin index, and the lattice spacing is taken to be unity.
The kinetic energy term describes hopping of electrons with spin $\alpha$ between neigbouring sites.
The hopping amplitudes are dimerized by the second term in $H$, given by 
\begin{eqnarray}
	H_\text{d} =  \sum_{n, \alpha}(-1)^{n}\Delta J (c^\dagger _{{n} +
		1,\alpha} c_{n,\alpha} + \text{h.c})\,.
\end{eqnarray}
%The dimerisation parameter $\Delta t_\mu$ is controlled via the application of an electric field $E_\mu$ and depends on the ferroelectric properties of the material (see Section ).
The next term is a staggered on-site potential 
\begin{eqnarray}
	H_\text{h} = h \sum_{n, \alpha, \beta}(-1)^{n}\sigma^z _{\alpha \beta} c^\dagger _{n,\alpha} c_{n,\beta}\,,
\end{eqnarray}
that couples antiferromagnetically to the spin.
Finally, the last term is given by 
\begin{eqnarray}
	H_\text{SO} =    \sum_{n, \alpha,\beta}i {\bm{\lambda}\cdot \bm{\sigma}}_{\alpha\beta}(c^\dagger _{n +
	1,\alpha} c_{n,\beta} - c^\dagger _{n ,\alpha} c_{n+1,\beta})\,,
\end{eqnarray}
where $|\bm{\lambda}|\ge 0$ is an arbitrary vector characterising the spin-orbit interaction and $\bm \sigma = \{\sigma^x,\sigma^y,\sigma^z\}$ are the Pauli matrices. 

The tight-binding model \eqref{eq:Ham 1D} is based on the antiferromagnetic spin-$\frac{1}{2}$ chain describing a class of crystalline materials that can be manipulated by external electromagnetic fields.
For example, metallic ferromagnetic compounds, such as Cu-benzoate~\cite{affleck1999field,haldane1983nonlinear} and $Yb_4As_3$~\cite{Kohgi01}, develop a staggered onsite potential when a perpendicular \textit{uniform} magnetic field is applied; as a consequence, they become insulating.
This is due to the Neel ground state induced by the competition between Dzyaloshinskii-Moriya interactions and a nonzero gyromagnetic tensor.
Additionally, ferroelectric materials, such $MnI_2$ and oxides $ABO_2$ with $A =  Cu$, $Ag$, $Li$ or $Na$ and $B=Cr$ or $Fe$, were shown to have an exchange interaction that depends on the applied electric field~\cite{xiang2011general,kimura2006inversion,seki2007impurity,seki2008spin}. 
The ferroic properties of such materials can be deployed in the design of experiments where electromagnetic fields act as control knobs to the system's parameters.

Motivated by the above discussion, we assume that the dimerization amplitude and staggered on-site potential depend on an external parameter $\phi$ through
\begin{eqnarray}
	\Delta J = \Delta J ^0 \cos(\phi)\text{   and   }h  = h_0 \sin(\phi)\,,
\end{eqnarray}
where $J^0$ and $h_0$ are constants.
Furthermore, we assume that spatio-temporal variations of $\phi$ are smooth enough such that the dynamics are well approximated by the first-order semiclassical equations, cf. Sec.~\ref{sec:Theory}.
The bulk spectrum of the Hamiltonian as a function of the external parameter $\phi$ is calculated by introducing periodic boundary conditions and applying Bloch's theorem to obtain the diagonalised Hamiltonian in terms of the quasi-momentum $k$.
In a suitable gauge, this is given by
\begin{eqnarray}
	H(k,\phi) = \bm{d}\cdot \bm{\gamma} + \bm{D}\cdot\bm{\Gamma}\,,
	\label{eq:1DHamMom}
\end{eqnarray}
where 
\begin{eqnarray}\begin{array}{rl}		
	\bm{d} =& \{J^+ + J^-\cos(k) ,J^-\sin(k),h\}\,,\\\\
	\bm{D} = &\{\lambda_x \sin(k) ,-\lambda_x +\lambda_x\cos(k) ,\lambda_y \sin(k) ,\\
	&\,\,\,-\lambda_y +\lambda_y \cos(k) ,\lambda_z \sin(k) ,-\lambda_z +\lambda_z \cos(k) \,\}\,,
	\end{array}
\end{eqnarray}
are real-valued vectors with $J^\pm = J\pm \Delta J$, 
\begin{eqnarray}
\begin{array}{c}
		\bm{\gamma} = \{\mathds{1}\otimes \sigma_x, \mathds{1}\otimes \sigma_y, \sigma_z\otimes \sigma_z\}	\,,
\end{array}
\end{eqnarray}
are three $4\times 4$ anticommuting unitary matrices $\{\gamma_i , \gamma_j\} = 2\delta_{ij}$, and, 
\begin{eqnarray}
	\begin{array}{rl}
		\bm{\Gamma} = &\{\sigma_x \otimes \sigma_x , \sigma_x \otimes \sigma_y ,\sigma_y \otimes \sigma_x ,\\
		& \,\,\,\sigma_y \otimes \sigma_y , \sigma_ z\otimes \sigma_x , \sigma_z \otimes \sigma_y\,\}\,,
	\end{array}
\end{eqnarray}
are six $4\times 4$ unitary matrices representing the spin-orbit interaction.

The bulk energy spectrum of the Bloch Hamiltonian \eqref{eq:1DHamMom} has four bands which can be intuitively described in the low energy limit by displaced Dirac-like cones in the $(k,\phi)$-parameter space, see Fig.~\ref{fig:fig1d}(b).
The ground state at half-filling is conducting only when $ h = \Delta J= 0$, while a nonzero dimerisation parameter $\Delta J$ or staggered potential $h$ induces a gap around zero energy. %~\footnote{These are the terms responsible for the induced gap in Refs.~\cite{}}.
The spectrum has a two-fold degeneracy in the entire BZ when $|\bm{\lambda}| = 0$, which is lifted to isolated points when $|\bm{\lambda}| > 0$.

The open boundary spectrum of the Hamiltonian is composed by the aforementioned bulk bands, in addition to two degenerate pairs of co-dimension 1 states, see Fig.~\ref{fig:fig1d}(c).	
Each pair disperses as a function of ${\phi}$ and merges into the bulk bands by crossing the gap. 
The probability distribution of these states is fully localized on the boundary with an exponential decay depending on the proximity to the bulk bands.
It is important to note that each pair has states localised at \textit{opposite} boundaries, hence, their combination induces a zero net charge polarization, see Fig.~\ref{fig:fig1d}(d).
On the other hand, the spin density associated to the operator $\hS = \sigma_z\otimes \mathds{1}$ exhibits spin localization at the boundaries that induces a nonzero spin polarization, see Fig.~\ref{fig:fig1d}(e).

Determining the symmetries of the system is crucial when characterizing the band structure.
For $h = 0$, the Hamiltonian~\eqref{eq:1DHamMom} has both time reversal (T.R.) symmetry $\Theta = i \mathds{K}\sigma_y \otimes \mathds{1}$ and chiral symmetry $\chi = \mathds{1}\otimes \sigma_z$ for any value of $\Delta J$.
On the other hand, the system has only T.R. symmetry in the entire phase-space, i.e.,
\begin{eqnarray}
	\Theta^{-1} H(k,\phi)\Theta = H(-k,-\phi)\,,
\end{eqnarray}
that is preserved for any value of $|\bm{\lambda}|$, $ h$, and $\Delta J$.
As such, the ground state can be decomposed into ``spin-sectors", namely
\begin{eqnarray}
\begin{array}{rl}
	\ket{\psi^1 ( k,\phi)}=& \Theta\ket{\psi^2 (- k,-\phi)}\\
	\ket{\psi^2 ( k,\phi)}= &-\Theta\ket{\psi^1 (- k,-\phi)}\,,
\end{array}
\label{eq:TRpartners}
\end{eqnarray}
formed by the T.R.-invariant partners in the $(k,\phi)$ parameter-space.
Classifying the topological properties of Hamiltonians depending on their dimensionality and symmetries has been well studied using various theoretical methods~\cite{bellissard1992gap, bellissard2001gap,shiozaki2017topological,kitaev2009periodic,chiu2016classification, ryu2010topological, altland1997nonstandard,qi2011topological}.
In our case, the parameter-space of the system provides an increased dimensionality that, in fact, can be ascribed a $\mathds{Z}_2$ index.
Below, we show that this topological invariant appears as a first-order correction to the spin transport and accumulation induced by an adiabatic drive, or weak inhomogeneities.

\subsubsection{Transport}\label{sec:1Dtrans}

%The transport of physical spin in one-dimension has been previously studied by Fu and Kane in the context of antiferromagnetic spin-chains~\cite{fu2006time}.
%It was shown that the periodic and adiabatic modulation of the system's parameters can, in fact, induce transport of physical spin across an otherwise insulating material.
%Here, we re-derive this result within the semiclassical framework.

The 1D Hamiltonian \eqref{eq:1DHamMom} can be adiabatically pumped by slowly changing the external parameter $\phi(t)$ over time, i.e., by temporally modulating the onsite energy and hopping terms in a periodic fashion. 
At each time $t$, the Hamiltonian is assumed to be diagonalized by a set of instantaneous bands, which now define a curvature tensor in phase-space $\Omega^{nm}_{k t}= i \dot{\phi}\left( \langle \partial_{\phi} n | \partial_{k} m \rangle  - \langle \partial_{k} n | \partial_{\phi}  m \rangle \right)$, where $(k,t)$ are the momentum-time coordinates and $\ket{n}$ denotes the set of occupied Bloch bands.
The spinor-current associated to an insulating ground state [cf. Eq.~\eqref{eq:current general}] is given by
\begin{eqnarray}
		j_{\hat{\mathcal{S}}} =  \frac{1}{2\pi	}\int\limits_{\mathds{T}^1 }  {\rm d k}\Tr {\underline{\mathcal{S}}} \Omega_{tk} \,,
	\label{eq:1dresult}
\end{eqnarray}
where $\hat{\mathcal{S}} = \mathds{1}\otimes \mathds{1}$ ($=\sigma_z \otimes \mathds{1}$) is the charge (physical spin) degree of freedom and the integration of the momentum $k$ is over a $\mathds{T}^1$ torus representing the entire 1D Brillouin zone.

When the time evolution is periodic, the spinor-charge transport after a pump period $T$, taken to be unity for simplicity, is equal to
\begin{eqnarray}
	\ds	\Delta q_{\hat{\mathcal{S}}}  &\ds= \int\limits_{0 } ^{T}  dt j_{\hat{\mathcal{S}}} 
	\eqcolon c_1\,.
	\label{eq:pumped charge}
\end{eqnarray}
Since the integration is over the closed momentum-time manifold, the transport of $\Delta q_{\hat{\mathcal{S}}}$ spinor-charges associated to $\hat{\mathcal{S}}$ is proportional to a \Fst spinor-Chern number $c_1\sim\int_{\mathds{T}^2} \rm d k\rm d\phi\Tr\underline{\mathcal{S}}\Omega_{k\phi}$ defined in the $(k,\phi)$-parameter space.

Following the modern approach to the definition of polarization, the bulk spinor-charge transport must be induced by the temporal gradient of the associated spinor-dipole moment density $\mathcal{P}^{\hat{\mathcal{S}}} $, i.e., $
	\Delta q_{\hat{\mathcal{S}}}    = \int_T \rm  dt \dot{\mathcal{P}}^{\hat{\mathcal{S}}} $.
For example, when $\hat{\mathcal{S}} = \mathds{1}\otimes \mathds{1}$ it corresponds exactly to the electric dipole moment, while for $\hat{\mathcal{S}}  = \sigma_z\otimes\mathds{1}$ it is the polarisation of physical spin~\cite{fu2006time}.
Comparing with Eq.~\eqref{eq:pumped charge}, the \Fst spinor-Chern number $c_1$ is identified with the contribution from a temporally modulated spinor-dipole moment density, i.e.,
\begin{eqnarray}
		c_1  \overset{!}{=}	\int\limits_T \rm  dt \dot{ \mathcal{P}}^{\hat{\mathcal{S}}} \,.
		\label{eq:dipole change is CN}
\end{eqnarray}
The above equality is a natural outcome of the semiclassical formalism: it relates a topological quantity in the system's parameter-space -- the \Fst spinor-Chern number $c_1$ -- to a macroscopic property of the material -- the rate of change of the spinor-dipole moment $\mathcal{P}^{\hat{\mathcal{S}}}$.

\begin{figure}
	\centering
	\includegraphics[width=1\linewidth]{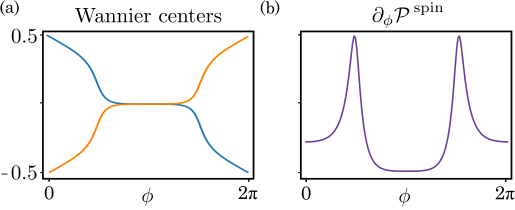}
	\caption{Wannier centres and polarization. (a) The Wannier centres of the two occupied bands as a function of the external parameter $\phi$, cf. Appendix~\ref{apx:poles}. 
	Each center ``winds" around the unit cell after each cycle. Crucially, the sum of the two vanishes as expected for the charge degree of freedom, i.e., when $\hat{\mathcal{S}} = \mathds{1}\otimes\mathds{1}$, but becomes nonzero for physical spin $\hat{\mathcal{S}} =\sigma_z\otimes \mathds{1}$.
(b) The derivative of the spin-dipole moment with respect to the external parameter $\phi$, in the case of $\hS = \sigma_z\otimes \mathds{1}$, i.e., for a physical spin.
The area under the graph is an integer equal to 1, and is related to the \Fst spin-Chern number of the $(k,\phi)$ parameter space, cf. Eq.~\eqref{eq:1D1CNtrans}.}
	\label{fig:fig1dpolcur}
\end{figure}

Focusing on the tight-binding model of Eq.~\eqref{eq:Ham 1D}, the \Fst spinor-Chern number associated to the charge degree, i.e., when $\hat{\mathcal{S}} = \mathds{1}\otimes  \mathds{1}$, is identically zero due to the trace property $\Tr( \mathds{1}\sigma_z) = 0$.
Indeed, the electric dipole moment of the Hamiltonian, shown in Fig.~\ref{fig:fig1dpolcur}(a), is decomposed into two contributions that originate from the two occupied Bloch states.
Each set of negative-energy (equivalently for positive-energy) bands can be written as a direct sum of two orthogonal eigenvectors that are mixed according to the strength of the spin-orbit interaction $\bm{\lambda}$. 
Importantly, T.R. symmetry allows for the decomposition of the polarization into contributions from the two spin sectors, cf.~Eq.~\eqref{eq:TRpartners}, where each sector ``winds" as a function of the external parameter $\phi$.
Hence, the net electric dipole moment vanishes, as it is given by the sum of two equal but opposite contributions; a consequence of T.R. symmetry.

On the other hand, the transport of physical spins after each cycle, calculated by taking $\hat{\mathcal{S}} = \sigma_z \otimes\mathds{1}$, becomes
\begin{eqnarray} c_1 = \frac{1}{4\pi	} \int_{\mathds{T}^2 } d\phi  {\rm d k} 
	\hat{\mathbf{\textbf{d}}}\cdot (\partial_{k}\hat{\textbf{\text{d}}} \times \partial_{\phi}\hat{\textbf{\text{d}}})=1\,,
	\label{eq:1D1CNtrans}
\end{eqnarray}
where $\hat{\bm d}= \bm d/|\bm d|$ and, now, the integration runs over a $\mathds{T}^2$ torus representing the momentum space and the periodic time evolution. %with period $T$, taken to be unity for simplicity.
Importantly, the first-order contributions to spin transport from a nonzero spin-orbit interaction $|\bm{\lambda}|$ vanish due to the trace properties of unitary matrices.
As a consequence, a quantized amount of physical spin, equal to $1$, will be transported after each cycle. 
From a macroscopic perspective, this is reflected in the spin-dipole moment density that acquires nonzero values and winds once as a function of $\phi$, see Fig.~\ref{fig:fig1dpolcur}(b).
Even though the spin-current of our model has a simple analytic solution, we note that for more complicated band structures, numerical tools can be used to calculate the relevant quantities from first-principles~\cite{king1993theory,benalcazar2017quantized,benalcazar2017electric}.

\subsubsection{Accumulation}\label{sec:1Dacc}
Before focusing on the spin degree of freedom, we briefly mention that the question of charge accumulation on a domain wall created by a spatially modulated 1D Hamiltonian has been previously studied in the context of high energy physics in Ref.~\cite{jackiw1976solitons}.
The system was shown to support a nontrivial solution, called soliton, which is exponentially localised on the domain wall and results in a fractional charge accumulation of $1/2$.
As we will see in this Section, the fractional accumulation of charge, and more generally of a spinor degree of freedom, is related to the \Fst spinor-Chern flux defined over the momentum-position coordinates of the system.

\begin{figure}
	\centering
	\includegraphics[width=1\linewidth]{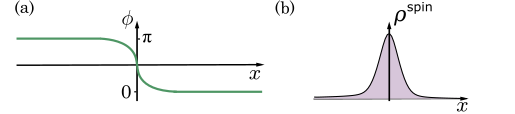}
	\caption{Domain wall. (a) The domain wall created by the parameter $\phi(r)$. (b) Sketch of the spin accumulation supported on the domain wall.}
	\label{fig:fig1ddw}
\end{figure}

In order to observe nontrivial effects in spin accumulation, the 1D model of Eq.~\eqref{eq:1DHamMom} is now modulated in position by introducing a spatial dependence in the external parameter $\phi(r)$, i.e., by changing the onsite energy and hopping terms in position space. 
Specifically, we assume that $\phi(r)$ acquires continuous values between $0$ and $\pi$, see Fig.~\ref{fig:fig1ddw}(a), with smooth enough modulations as compared to the size of the wavepacket. 
In this regime, the curvature tensor of the local bands is given by $\Omega^{nm}_{k r}= i \partial_r {\phi}\left( \langle \partial_{\phi} n | \partial_{k} m \rangle  - \langle \partial_{k} n | \partial_{\phi}  m \rangle \right)$ and the induced spinor-charge density at half-filling, cf. Equation~\eqref{eq:DoS general}, by
\begin{eqnarray}
	\rho_{\hat{\mathcal{S}}} =  \frac{1}{2\pi	}\int\limits_{\mathds{T}^1 }  {\rm d k}\Tr\underline{{{\mathcal{S}}}} \Omega_{rk} \,,
\end{eqnarray}
where $(k,r)$ are the momentum-position coordinates, and the integration of the momentum $k$ is over the entire 1D Brillouin zone.
The total spinor-charge accumulation $q_{\hat{\mathcal{S}}}$ in a region $C$ enclosing the domain wall is, thus, given by 
\begin{eqnarray}
			q_{\hat{\mathcal{S}}}   &\ds= \int\limits_{C}   \rm dr \rho_{\hat{\mathcal{S}}}\eqcolon \Phi_1\,,
	\label{eq:1d spin accum}
\end{eqnarray}
where $\Phi_1\sim\int_{\left[0,\pi\right]\times \mathds{T}} \rm d\phi\rm d k\Tr\underline{\mathcal{S}}\Omega_{k\phi}$ is the \Fst spinor-Chern flux attached to the 2D Dirac-like cones in the $(k,\phi)$ parameter space, cf. Fig.~\ref{fig:fig1d}(b).
We emphasize that Eq.~\eqref{eq:1d spin accum} is valid for any insulating ground state and is not limited to the Hamiltonian~\eqref{eq:1DHamMom}.

Within the classical approach of multipole moments, Eq.~\eqref{eq:1d spin accum} can be alternatively described by the spatial gradient of a bulk spinor-dipole density, i.e., $
	q_{\hat{\mathcal{S}}} =  \int_{{C}} \rm dr	\partial_r\mathcal{P}^{\hat{\mathcal{S}}}\,$, where $\mathcal{P}^{\hat{\mathcal{S}}}$ is again the spinor-dipole moment density and $C$ is the integration domain over space (related to an integration over $\phi$ by the appropriate coordinate transformation).
Consequently, the \Fst spinor-Chern flux $\Phi_1$ is identified with the contribution from a spatially modulated spinor-dipole moment density, i.e.,
\begin{eqnarray}
	\Phi_1\overset{!}{=}  \int\limits_C \rm  dr \partial_r \mathcal{P}^{\hat{\mathcal{S}}}\,.
\end{eqnarray}
%where integration domain $C$ does not enclose the boundary of the material.
Similar to the derivation of spinor-charge transport, the above equation leads to a fundamental connection between an abstract geometrical quantity and an electronic property of the material.

In general, the integration domain $C$ in Eq.~\eqref{eq:1d spin accum} does not necessarily cover a closed manifold in the	 $(k, \phi)$-parameter space, therefore, $\Phi_1$ is not expected to be quantized.
However, under symmetry constraints the \Fst spin-Chern flux of the tight-binding model of Eq.~\eqref{eq:1DHamMom} can, in fact, become quantized and lead to a fractional spin accumulation localised at the domain wall.
Specifically, in its chiral limit, i.e., when $h_0\to 0$, the \Fst spin-Chern flux associated to the spin operator $\hat{\mathcal{S}} = \sigma_z\otimes \mathds{1}$ becomes quantized to
\begin{eqnarray}
	\Phi_1 
		 \ds= \frac{1}{4\pi	} \int_{ \mathds{T}\times\left[0, \pi\right] }^{  }  {\rm d k}\rm{d}\phi \,
		\hat{\mathbf{\textbf{d}}}\cdot (\partial_{k}\hat{\textbf{\text{d}}} \times \partial_{\phi}\hat{\textbf{\text{d}}})\overset{h_0\to0^+}{=}\frac{1}{2}\,,
	\label{eq:1stspinflux}
\end{eqnarray}
where $ \mathds{T}\times\left[0, \pi\right] $ represents the one-dimensional Brillouin zone $\mathds{T}$ and the integration region $\left[0, \pi\right]$ in the $\phi$-parameter space; the total phase-space volume is taken to be unity for simplicity.
As a result, the ground state of the material exhibits localisation of spin on the domain wall created by the spatial modulation of the Hamiltonian, see Fig.~\ref{fig:fig1ddw}(b).
Importantly, the amount of spin that is supported on the domain wall is \textit{quantized} and equal to $1/2$.
Indeed, the bulk spin-dipole moment of each spin sector changes by a fractional amount between $\phi = 0$ and $\pi$, cf. Fig.~\ref{fig:fig1dpolcur}~(a), as a consequence of chiral symmetry $\chi$.
% Thus, the toatal accumulation of physical spin at the interface defined by the domain wall is equal to $1/2$.

\subsection{In two dimensions}\label{sec:2D}
In this Section, we use the semiclassical theory to establish 
a connection between quantised spinor-charge transport (accumulation) in 2D insulators and the \Snd spinor-Chern number (flux) characterizing the system's phase-space.
We illustrate this connection by calculating the spinor-current up to second order in the adiabatic driving of a concrete tight-binding model describing spinful electrons in a square lattice with nearest neighbor coupling.
In particular, we show that the transported spin after a pump cycle is proportional to not just the \Fst sub-spin-Chern number (discussed in Sec.~\ref{sec:1D}), but also to a \Snd spin-Chern number defined over the four-dimensional parameter-space of the system.
Similarly, the spin-density is related to the \Fst and \Snd spin-Chern fluxes that give rise to fractional spin accumulation at the edges and corners, respectively, when global symmetry constraints are imposed.
%We show that for the charge degree of freedom, such quantities vanish due to T.R. symmetry.
Finally, we decompose the spin transport and accumulation in terms of modulations of spin-multipole moments and propose a dynamical scenario where an adiabatically driven, weakly inhomogeneous 2D crystal exhibits a quantised spin transport and fractional spin accumulation with co-dimensions 2.

\begin{figure}[h!]
	\centering
	\includegraphics[width=1\linewidth]{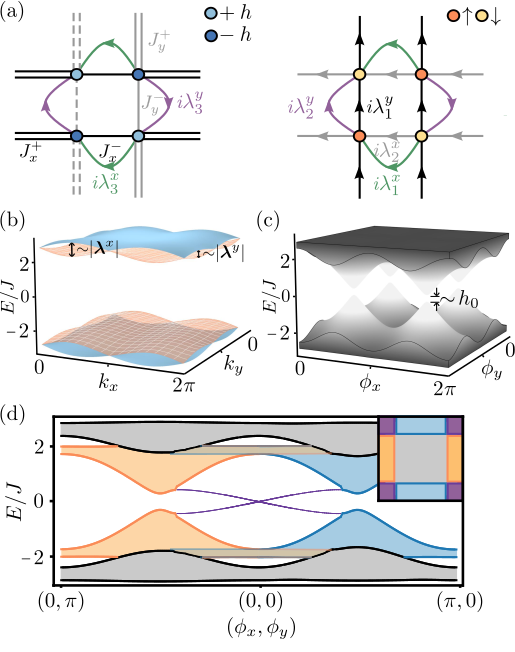}
	\caption{The model Hamiltonian in two dimensions. 
		(a) \textit{left}: The tight-binding lattice described by $H$, cf. Eqs.~\eqref{eq:2DHamKE}-\eqref{eq:2DHamSO}.
		Electrons with spin $\alpha$ tunnel to the nearest neighbour in the $\mu$-direction with amplitude $J_\mu ^\pm = J_\mu \pm \Delta J_\mu$, denoted by single/double lines. 
		The light/dark blue sites denote an onsite potential with positive/negative sign depending on $\alpha$.
		The $\sigma_z$ component of the spin-orbit interaction in the different directions is denoted by the green/purple curved lines.
		\textit{right}: The tight-binding lattice describing hopping between spin up and down electrons, denoted by the yellow and orange sites, respectively.
		The $\sigma_x$ ($\sigma_y$) component of the spin-orbit interaction in the different directions is denoted by the green (gray) and black (purple) lines.
		(b) The bulk spectrum of the Hamiltonian $H(\bm k, \bm \phi)$ at $\bm \phi = (0,0)$ showing four doubly-degenerate bands.
		(c) The bulk spectrum as a function of the external parameters showing the four 4D Dirac-like crossings where the mass term is proportional to $h_0$.
		Each point in the $\bm \phi$ plane represents the bulk spectrum in the entire Brillouin zone $\bm k$, projected onto the perpendicular axis.
		(d) The open boundary spectrum of the Hamiltonian showing bulk (gray), edge (blue and orange), and corner (purple) states.
		The inset shows a schematic representation of the different states found in the open boundary spectrum.
		For all simulations we used $J _\mu/10= J_\mu - \Delta J^0 _\mu =0.1$. 
		In (b) we used $\lambda^x _1 /6 = \lambda^y _2 /3 = 0.1$ to emphasize the effect of spin-orbit coupling in the different directions.
		In (c) and (d) we used $\lambda^x _1  = \lambda^y _2 = 0.1$ where the former has $h_0 =0.1$ and the latter $h_0 =0.5$.
		}
	\label{fig:fig2d}
\end{figure}

A tight-binding model that contains all necessary ingredients is shown in Fig.~\ref{fig:fig2d}(a).
It describes non-interacting spinful electrons on a 2D square lattice with Hamiltonian $H = H_\text{\text{KE}} + H_\text{d} + H_\text{h} +  H_\text{SO} $.
The first term is the kinetic energy, given by 
\begin{eqnarray}
    H_\text{\text{KE}} =  \sum_{\bm{n},\mu, \alpha}J_\mu(e^{iA_\mu}c^\dagger _{{\bm{n}} +	\hat{\mu},\alpha} c_{\bm{n},\alpha} + \text{h.c})\,,
    \label{eq:2DHamKE}
\end{eqnarray}
where $J_\mu$ is the hopping amplitude in the $\mu$-th direction, $\bm n=(n_x, n_y)$ is the position vector in the 2D lattice, $\hat{\mu}$ is the unit vector in the $\mu$ direction, $\alpha$ the spin index, $\bm A= \left(0,\pi n_x\right)$ is a static vector potential and the lattice spacing is taken to be unity.
The kinetic term describes hopping of electrons with spin $\alpha$ between neigbouring sites on a square lattice with a $\pi$-flux quanta per plaquette.
Similar to the one-dimensional case of Eq.~\eqref{eq:Ham 1D}, the second term in $H$ defines the dimerisation of the hopping amplitudes in the two directions 
\begin{eqnarray}
    H_\text{d} =  \sum_{\bm{n},\mu, \alpha}(-1)^{n_\mu}\Delta J_\mu (e^{iA_\mu}c^\dagger _{{\bm{n}} +
	\hat{\mu},\alpha} c_{\bm{n},\alpha} + \text{h.c})\,.
\end{eqnarray}
The next term is a checkerboard on-site potential $H_\text{h} = h \sum_{\bm{n}, \alpha, \beta}(-1)^{n_x + n_y}\sigma^z _{\alpha \beta} c^\dagger _{\bm{n},\alpha} c_{\bm{n},\beta}$ that couples antiferromagnetically to the physical spin, while the last term is given by 
\begin{eqnarray}
    H_\text{SO} =    \sum_{\bm{n},\mu, \alpha,\beta}i {\bm \lambda^\mu \cdot \bm{\sigma}}_{\alpha\beta}e^{iA_\mu}(c^\dagger _{\bm{n} +
	\hat{\mu},\alpha} c_{\bm{n},\beta} - c^\dagger _{\bm{n} ,\alpha} c_{\bm{n}+
	\hat{\mu},\beta})\,,\notag\\
    \label{eq:2DHamSO}
\end{eqnarray}
where $\bm{\lambda}^x = \{\lambda^x_1,\lambda^x_2,\lambda^x_3\}$ is an arbitrary vector characterising the spin-orbit interaction along the $x$-direction (respectively for the $y$-direction) and it plays a role similar to Dzyaloshinskii-Moriya interaction. 

%%%%%%%%%%%%%%%%%%%%%%%%%%%%%%%%%%%%%%%%%%%%%%%%%%%%%%%%%%%%
Our tight-binding modelling is motivated by recent experiments in 2D magnetic $SrIrO_3$ and $SrTiO_3$ materials~\cite{hao2018giant} where a highly efficient control of the antiferromagnetic order was demonstrated using a uniform magnetic field.
Alternative implementations can also be found in piezo-electric and piezo-magnetic crystals, where electric and magnetic properties are controlled by lattice deformations.
With these studies in mind, we assume that the dimerisation parameter and onsite checkerboard potential depend on two external parameters $\bm{\phi} = (\phi_x , \phi_y)$,
\begin{eqnarray}
	\Delta J_\mu = \Delta J_\mu ^0 \cos(\phi_\mu)\text{   and   }h  = h_0 \prod_\mu\sin(\phi_\mu)\,,
\end{eqnarray}
where the spatio-temporal variations of $\bm \phi$ are smooth enough such that the system can be expanded in terms of a local Hamiltonian and the semiclassical dynamics are well-captured within second-order perturbation theory, cf. Sec.~\ref{sec:Theory}.% and Appendix~\ref{apx:semi}.

The momentum space Hamiltonian as a function of the external parameters $\bm \phi$ is given by
\begin{eqnarray}
	H(\bm k,\bm\phi) = \bm{d}\cdot \bm{\gamma} + \bm{D}\cdot\bm{\Gamma}\,,
	\label{eq:2DHam PS}
\end{eqnarray}
where, now, $\bm d$ and $\bm D$ are 5- and 12-vectors, respectively,
\begin{eqnarray}\begin{array}{rl}		
		\bm{d} =& \{J^+ _x + J^-_x\cos(k_x) ,J^-_x\sin(k_x),\\
		&\,\,\,	J^+ _y + J^-_y\cos(k_y) , J^-_y\sin(k_y),h\}\,,\\\\
		\bm{D} = &\{\lambda^x_1 \sin(k_x) ,-\lambda^x_1 +\lambda^x_1\cos(k_x) ,\lambda^x_2 \sin(k_x) ,\\
		&\,\,\,-\lambda^x_2 +\lambda^x_2 \cos(k_x) ,\lambda^x_3 \sin(k_x) ,-\lambda^x_3 +\lambda^x_3 \cos(k_x) ,\\
		&\,\,\,\lambda^y_1 \sin(k_y) ,-\lambda^y_1 +\lambda^y_1 \cos(k_y) ,\lambda^y_2 \sin(k_y),\\
		&\,\,\, -\lambda^y_2 +\lambda^y_2 \cos(k_y) ,\lambda^y_3 \sin(k_y) ,-\lambda^y_3 +\lambda^y_3 \cos(k_y) \,\}\,,
	\end{array}\notag
\end{eqnarray}
with $J^\pm _\mu = J_\mu\pm \Delta J_\mu$, and
\begin{eqnarray}
	\begin{array}{rl}
		\bm{\gamma} = &\{\mathds{1}\otimes \sigma_x\otimes \mathds{1}, \mathds{1}\otimes  \sigma_y\otimes \sigma_z,\\
		&\,\,\,  \mathds{1}\otimes  \sigma_y\otimes \sigma_y, \mathds{1}\otimes \sigma_y\otimes \sigma_x, \sigma_z\otimes\sigma_z\otimes \mathds{1}\}\,,	
	\end{array}\notag
\end{eqnarray}
are five $8\times 8$ anticommuting hermitian matrices $\{\gamma_i , \gamma_j\} = 2\delta_{ij}$. Lastly, the spin-orbit interaction is represented by twelve $8\times 8$ unitary matrices
\begin{eqnarray}
	\begin{array}{rl}
		\bm{\Gamma} = &\{\sigma_x\otimes \sigma_x\otimes \mathds{1}\,\,,
		\sigma_x\otimes  \sigma_y\otimes \sigma_z,
		\sigma_y\otimes \sigma_x\otimes \mathds{1},\\
		& \,\,\,\sigma_y\otimes  \sigma_y\otimes \sigma_z,
		\sigma_z\otimes \sigma_x\otimes \mathds{1}\,\,,
		\sigma_z\otimes  \sigma_y\otimes \sigma_z,\\
		& \,\,\,
		\sigma_x\otimes  \sigma_y\otimes \sigma_y,
		\sigma_x\otimes  \sigma_y\otimes \sigma_x,
		\sigma_y\otimes  \sigma_y\otimes \sigma_y,\\
		& \,\,\,
		\sigma_y\otimes  \sigma_y\otimes \sigma_x,
		\sigma_z\otimes  \sigma_y\otimes \sigma_y,
		\sigma_z\otimes  \sigma_y\otimes \sigma_x\,\}\,.
	\end{array}\notag
\end{eqnarray}

The energy spectrum of the Hamiltonian~\eqref{eq:2DHam PS} has eight bands, see Fig.~\ref{fig:fig2d}(b).
Each set of positive- and negative-energy bands can be written as a direct sum of two orthogonal groups of eigenvectors that are coupled by the spin-orbit interaction.
%These can be qualitatively grouped into four pairs where each pair is composed by time-reversal invariant partners.
Each pair of groups remains degenerate in the entire Brillouin zone when the spin-orbit interaction vanishes $|\bm \lambda^\mu| = 0$, otherwise, the degeneracy survives only at isolated points.
At half filling the system is conducting only when $ h = \Delta J_\mu = 0$, and insulating if either the staggered potential $h$ or the dimerisation parameters $\Delta J_x$ or $\Delta J_y$ become nonzero, see Fig.~\ref{fig:fig2d}(c).

Introducing open boundary conditions and solving for the eigenenergies we find that in addition to the bulk bands, the spectrum has two sets of co-dimension 1 (edge) states: (i) right/left localised states and (ii) top/bottom localised states, see Figs.~\ref{fig:fig2d} (d).
As a function of $\bm{\phi}$, the edge states disperse and merge into the bulk bands without crossing the gap.
Each set of right/left or top/bottom states is localised in opposite sites of the crystal, hence, their sum vanishes when calculating the charge polarization of the ground state at half-filling.
Similarly, the spin density distribution of edge states has vanishing contribution to the total spin polarization.

The spectrum supports an additional set of co-dimension 2 states localised at the corners.
In contrast to the co-dimension 1 states, as a function of $\bm{\phi}$ the corner states disperse, merge with edge or bulk states, and most importantly cross the gap.
However, such spectral flow does not induce any net charge transport since the states that cross the gap are made up of two electrons and two holes; on the other hand, a heat and spin transport is expected to show nontrivial effects, as we show below.

The Hamiltonian~\eqref{eq:2DHam PS} has T.R. symmetry $\Theta = i\mathds{K} \sigma_y \otimes \mathds{1} \otimes \mathds{1} $ and chiral symmetry $\chi = \mathds{1} \otimes\sigma_z\otimes  \mathds{1} $ for any value of $\Delta J_\mu$ and $\bm \lambda^\mu$, only when $h = 0$. 
Additionally, the Hamiltonian has charge conjugation symmetry $C=i\mathds{K} \sigma_y \otimes \sigma_z \otimes \mathds{1} $ in the entire phase-space, i.e.,
\begin{eqnarray}
    C^{-1}H(\bm k, \bm \phi)C = -H(-\bm k , - \bm \phi)\,,
\end{eqnarray}
for any value of $\Delta J_\mu$, $\bm \lambda^\mu$, and, $h $.
Hence, the occupied subspace of Bloch states can be partitioned into conjugate partners with a characteristic 4D $Z_2$ index~\cite{altland1997nonstandard}.
As we show below, this index emerges in the second order corrections to the spin transport and accumulation that is induced by spatio-temporal modulations of the crystal.

As the dimensionality of phase-space can now support a variety of nontrivial curvatures, here we summarize their physical origins.
As discussed in Sec.~\ref{sec:1D}, adiabatic drives give rise to mixed momentum-time curvatures $\Omega_{k_i t}$, while weak inhomogeneities give rise to momentum-position curvatures $\Omega_{k_i r_j}$.
Any weak external magnetic field that threads the insulator is incorporated via the position curvature $\Omega_{r_i r_j}\equiv B_{ij}$, while electric fields via the mixed position-time curvature $\Omega_{t r_i}\equiv E_i$.
Lastly, a momentum curvature $\Omega_{k_i k_j}$ arises as the relevant geometrical quantity in momentum space.

\subsubsection{Transport}\label{sec:2Dtrans}

The topological corrections to electronic charge transport up to second order have been extensively studied using the semiclassical theory~\cite{Xiao2010,  Gao2015, Price2015, price2016measurement}.
At first-order, Hall currents were shown to be related to a \Fst Chern number in momentum space, while second-order effects gave rise to 2D topological charge pumps with a \Snd Chern number response.
The relation of these indices to robust boundary physics, namely, to co-dimension 1 and 2 states, has been experimentally demonstrated in cold atomic clouds~\cite{Lohse2018,Lohse2016,Nakajima2016,Goldman:2016NatPhys,Goldman:2016NatPhys, cooper2018,mei2019topological}, photonic lattices~\cite{Kraus2012a,Verbin2015,lu2016topological,khanikaev2017two,ozawa2019topological,benalcazar2020higher,Zilberberg2018}, metamaterials~\cite{serra2018observation,apigo2018topological,cheng2020demonstration,long2019floquet,xia2020experimental,grinberg2020robust,rosa2019edge,tsai2019topological}, and electrical circuits~\cite{liu2019topologically,peterson2018quantized,imhof2018topolectrical,serra2019observation,ezawa2019electric,yu20194d}.

Here, we extend this description to spin degrees and derive the well-known spin-Hall effect, as well as, novel 2D topological spin pumps.
Specifically, we show how a \Fst and \Snd spin-Chern number manifest as corrections to the spin-current and highlight the relation to the quantised changes of spin-multipole moments.

The semiclassical equations of motion~\eqref{eq:EoM nonlinear} valid up to second order in perturbation theory result in a nonvanishing spinor-current,
\begin{eqnarray}
	j^i _{\hat{\mathcal{S}}} =  \int\limits_{\mathds{T}^2 } \frac{\rm d^2 k}{(2\pi)^2}\Tr\underline{S}\Big(\Omega_{t k_i} +\epsilon^{ij}\epsilon^{lm} \Omega_{t k_l}\Omega_{r_j k_m}\Big.\notag\\
	\hspace{70pt}\Big.+\epsilon^{ij}\Omega_{k_x k_y}E_j \Big)\,,
	\label{eq:2dresult}
\end{eqnarray}
where the integration domain is over a $\mathds{T}^2$ torus representing the Brillouin zone, Latin indices run over two spatial coordinates $i \in\{x,y\}$, and $\underline{S}$ is the matrix representation of the spinor operator in the basis of occupied states.
% with $\hat{\mathcal{S}}=\mathds{1}\otimes\mathds{1}\otimes\mathds{1}$ (or $\mathds{1}\otimes\sigma_z\otimes\mathds{1}$) the charge (spin) operator. 
The spinor-current has contributions from: (i) a curvature in $(t,k_i)$ coordinate space representing the adiabatic drive of the $\Delta J_i$ dimerization parameter, (ii) a product of curvatures in the $(\bs k, \bm r , t)$ coordinate space from the simultaneous drive and deformation of the crystal, and (iii) a combination of momentum curvature and applied electric field. 
When $\hat{\mathcal{S}}$ represents the charge degree, the latter is reduced to the usual quantum Hall response where the current depends on the \Fst Chern number in momentum space $c_1 \sim \int_{\mathds{T}^2 } {\rm d^2 k}\Omega_{k_x k_y}$.
Analogously, when $\hat{\mathcal{S}}$ is the spin degree we obtain the quantum spin-Hall effect with a spin current response proportional to a $\mathds{Z}_2$ topological index $c_1 \sim \int_{\mathds{T}^2 } {\rm d^2 k}\uS\Omega_{k_x k_y}$.
As these effects are well established, we neglect them for the remaining calculations.

The transport of spinor-charges in the $i$-th direction after a pump cycle $t\in\left[t_0,t_0 + T\right]$ is given by
\begin{eqnarray}
	\Delta q_{\hS}^i   =  \int\limits_{T,V }   \text{d} t \text{d}^2  r \,j^i _{\hat{\mathcal{S}}} = c_1 ^{i}+ c^i _2\,,
	\label{eq:2d pumped charge}
\end{eqnarray}
where the integral runs over a period $T$ and the volume of the unit cell $V\in\left[0,l_x\right]\times \left[0,l_y\right]$, hereafter, taken to be unity for simplicity.
Remarkably, we find that the perturbative corrections to the spinor transport are proportional to topological quantities defined in the $\bm\xi = (\bm k , \bm \phi)$ parameter space.
Namely, at first order we find a \Fst sub-spinor-Chern number $c^{i}_1$, defined as
%\begin{eqnarray}
%	c^{tk_i}_1 = \frac{1}{(2\pi)^2V} \int\limits_{T,V ,\mathds{T}^2 }   \rm dt\rm d^2r \rm d^2 k \Tr\underline{S} \Omega_{t k_i}\,,
%\end{eqnarray}
\begin{eqnarray}
	c^{i}_1 = \frac{1}{(2\pi)^2} \int\limits_{\mathds{T}^4 } |\mathcal{J}_j|  \rm d^4  \xi \Tr\underline{S} \Omega_{\phi_j k_i}\,,
\end{eqnarray}
where $|\mathcal{J}_j| = \dot{\phi}_j$ is the Jacobian of the transformation between the time coordinate and $\bm \phi$.
We note that in the 1D case, cf.  Eq.~\eqref{eq:1D1CNtrans}, the Jacobian is unity as it is given by the transformation of a single normalized coordinate.
At second order, a \Snd spinor-Chern number arises, defined as
%\begin{eqnarray}
%	c^i_2 = \frac{1}{(2\pi)^2V} \int\limits_{T,V ,\mathds{T}^2 }   \rm dt\rm d^2r \rm d^2 k \epsilon^{ij}\epsilon^{lm}\Tr\underline{S} \Omega_{t k_l}\Omega_{r_j k_m}\,. 
%\end{eqnarray}
\begin{eqnarray}
	c^i_2 = \frac{1}{(2\pi)^2} \int\limits_{\mathds{T}^4 } |\mathcal{J}_{jnm}|   \rm d^4  \xi \epsilon^{ij}\epsilon^{ls}\Tr\underline{S} \Omega_{\phi_n k_l}\Omega_{\phi_m k_s}\,,
	\label{eq:2ndCNtrans}
\end{eqnarray}
where $|\mathcal{J}_{jnm}| =\dot{\phi}_n \partial_{r_j}\phi_m$ is the Jacobian of the transformation between position-time coordinates and $\bm \phi$. 
Note that Eq.~\eqref{eq:2ndCNtrans} indeed defines the set of \Snd spinor-Chern numbers that characterise the momentum-position-time coordinate space [cf.~Eq.\eqref{eq:second_chern}] since terms proportional to $\Omega_{t r_j}$ will vanish in the absence of external electromagnetic field perturbations.

Extending the electric polarization in macroscopic materials to higher spinor-multipole moments, we decompose the bulk spinor-current as
\begin{eqnarray}
	j^{i}_{\hS } =   - \partial_{t} \mathcal{P}^{\hS} _{i} + \partial_{t} \partial_{r_j} \mathcal{Q}^{\hat{\mathcal{S}}}_{ij}\,,
	\label{eq:multi curent 2}
\end{eqnarray}
where $\bm{\mathcal{P}}^{\hS}$ is the spinor-dipole moment density vector, and, $\mathcal{Q}_{ij}^{\hS}$ is the spinor-quadrupole moment density in position space.
% The terms in the parenthesis depend on the particular details of the boundary and are summarized in Appendix~\ref{}.
Equations~\eqref{eq:2dresult} and~\eqref{eq:multi curent 2} establish a fundamental connection between topological quantities defined in the system's phase-space and the modulations of the spinor-dipole and spinor-quadrupole moment densities, namely,
\begin{eqnarray}
	c^{i}_1+ c^i_2\overset{!}{=} \int\limits_{T,V  }   \rm dt\rm d^2r   \left( -\partial_{t} {\mathcal{P}^{\hS}}_i + \partial_{t} \partial_{r_j} \mathcal{Q}^{\hat{\mathcal{S}}}_{ij}\right)\,.
	\label{eq:geomul2}
\end{eqnarray}
We emphasize that the above equation is independent of the particular Hamiltonian and can be used as a general geometrical definition of the spinor-multipole moments in two dimensions; a definition that eliminates any gauge ambiguity, as it is based on integrated differences.

Focusing on our specific tight-binding model of Eq.~\eqref{eq:2DHam PS}, nontrivial spinor-currents can be induced by adiabatically driving and periodically modulating in space the external parameters $\bm \phi$.
For simplicity, we assume $\dot{\phi}_x = 2\pi/T$ and $\partial_x \phi_y = 2\pi/{l}$ with $T$ (${l}$) the period (lengthscale) of the modulation.
The proper static deformation of the internal parameters is a crucial ingredient in 2D topological pumps as without it the \Snd spinor-Chern number becomes trivial, see for example Ref.~\cite{kraus2013}.
By defining $\hat{\mathcal{S}}=\mathds{1}\otimes\mathds{1}\otimes\mathds{1}$ (or $\sigma_z\otimes\mathds{1}\otimes\mathds{1}$) as the charge (spin) operator, we calculate the \Fst sub-spinor-Chern numbers and find that they identically vanish for both the charge and spin degree of freedom, i.e., $c_1^i = 0$.
Additionally, the \Snd Chern number associated to the charge degree of freedom is zero because of the trace properties of hermitian matrices.
On the other hand, the \Snd spin-Chern number associated with the physical spin $\hat{S}=\sigma_z\otimes\mathds{1}\otimes\mathds{1}$ becomes nontrivial and equal to \begin{eqnarray}
    c^i_2 =  \frac{3}{8\pi^2}\int_{\mathds{T}^4}\epsilon^{ij}\delta_{j,y} \mathrm{d}^4 \xi\hat{\mathbf{\textbf{d}}}\cdot (\partial_{\mathrm{k}_x}\hat{\textbf{\text{d}}} \times \partial_{\mathrm{k}_y}\hat{\mathbf{\textbf{d}}}\times \partial_{\phi_x}\hat{\mathbf{\textbf{d}}} \times \partial_{\phi_y}\hat{\mathbf{\textbf{d}}})\notag\,,\\
    \label{eq:2ndCNTB}
\end{eqnarray}
where $\hat{\bm d}= \bm d/|\bm d|$, and, $\delta_{i,j}$ is the Dirac delta function.
The latter is a consequence of the chosen driving scheme and stems from the Jacobian transformation in Eq.~\eqref{eq:2ndCNtrans}.
Hence, spin transport can be readily understood in the framework of a 4D $\mathds{Z}_2$ insulator where the top topological invariant is given by the difference of the ``mirror" \Snd Chern numbers characterizing the eigenstates from the two spin sectors.

\begin{figure}
	\centering
	\includegraphics[width=1\linewidth]{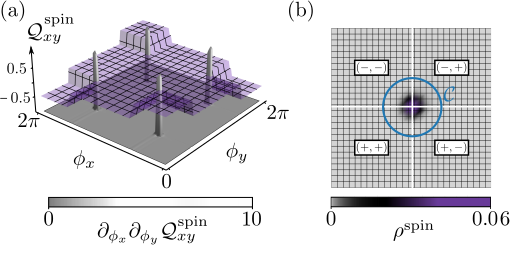}
	\caption{Spin observables in the chiral limit. (a) The spin-quadrupole moment (cf. Appendix~\ref{apx:poles}) ``winds" twice in the two-dimensional parameter space $\bm \phi$.
	This is in correspondence with the four singular contributions to the \Snd spin-Chern number, where each integrates to $1/2$.
% 	The spin-quadrupole moment (cf. Appendix~\ref{apx:poles}) as a function of $\bm \phi$ and the corresponding \Snd spin-Chern flux.
% 	The spin-quadrupole moment ``winds" twice in the two-dimensional parameter space, as indicated by the sigularities in the \Snd spin-Chern flux where each integrates to $1/2$.
	(b) The spin density of the electronic ground state at half filling, cf. Eqs.~\eqref{eq:2dDos} and~\eqref{eq:2dClassicalPoles}.
	The $\pm$ sign shown in the four quadrants indicates the values of $\bm \phi$ ($0$ or $\pi$, respectively) and creates a domain wall that supports a nonzero spin density.
	The total amount of spin in region $\mathcal{C}$ is quantized and equal to $1/2$, cf. Eq.~\eqref{eq:2dSpinQuant}. 
		For the simulations we used $J_\mu = 1$, $\Delta J^0 _\mu= 0.9$, $|\bm\lambda^\mu|  = \lambda^\mu_1= 0.1$ (for both $\mu = x$ and $y$), and $h_0 = 0.001$.
		}
	\label{fig:fig2dpoles}
\end{figure}

The relation between spin transport, the \Snd spin-Chern number, and the spin-quadrupole moment is illustrated in Fig.~\ref{fig:fig2dpoles}(a).
First, we note that the calculated electric- and spin-dipole moments vanish in the entire phase-space, reflecting the trivial value of the \Fst spinor-Chern number. 
Similarly, the electric quadrupole moment is decomposed into two equivalent contributions that originate from the two occupied (doubly degenerate) spin sectors.
As these contributions come with opposite signs, the net electric quadrupole moment is zero in the entire parameter space; this is a consequence of the trivial value of the \Snd Chern number associated to the charge degree of freedom.
On the contrary, the spin-quadrupole moment of the model becomes nonzero and, in fact, ``winds" twice as a function of the external parameters $\bm\phi$.
This higher-dimensional winding manifests as a nontrivial \Snd spin-Chern number, given by Eq.~\eqref{eq:2ndCNTB}, and to a quantized spin transport averaged over a pump cycle.

\subsubsection{Accumulation}\label{sec:2Dacc}

%Before focusing on spin accumulation we note the immense progress that has been done for the charge degree of freedom.
In a two-dimensional system there are two kinds of boundary states that can appear: with co-dimensions 1 or 2.
The former corresponds to states localized in one direction but extended in the other; found, for example, on the edges of Hall systems or in insulators with nonzero intrinsic polarization.
On the other hand, the interesting properties of co-dimension 2 states have only recently been rigorously explored.
This lead to the prediction and observation of states which are localized in both dimensions and, under certain symmetry constraints, carry a quantized charge $\pm 1/2$~\cite{Zilberberg2018,serra2018observation,imhof2018topolectrical,ryu2009masses,lin2017topological,hashimoto2017edge,langbehn2017reflection,benalcazar2017quantized,trifunovic2018higher,geier2018second,schindler2018higher,wang2018higher,ezawa2018minimal,ghorashi2019second,cualuguaru2019higher,teo2010topological}.
From these studies, a new class of TIs emerged, dubbed ``higher-order TIs", where a $d$-dimensional insulator has nontrivial boundary phenomena manifesting at its $d-h$ boundary, where $h\ge 1$.
The associated electric multipole moments of higher-order TIs can be readily calculated using the modern theory of Wilson, and nested-Wilson loops~\cite{benalcazar2017quantized}.
For co-dimension 2 states, the key observable is the quadrupole moment which is constrained to obtain only certain values and, as a result, quantize the accumulation of electronic charge at the corner.

As we will see in this Section, an alternative definition of second-order TIs and its extension to, what we dub, ``spin second-order TIs" is naturally obtained within the semiclassical theory.
Specifically, we show that geometrical properties of phase-space -- the spin-Chern fluxes -- appear as corrections to the spin density, directly leading to a fractional accumulation at the 0D or 1D boundaries when symmetry constraints are imposed.
Finally, we show how these quantities are related to the spin-multipole moments, namely, the spin-dipole and spin-quadrupole moment. %; this reveals a fundamental connection between topology of phase-space and higher-order TIs. 

In general, a weakly inhomogeneous insulator in two spatial dimensions is well-characterized by the spinor-charge density up to second-order corrections [cf.~Eq.~\eqref{eq:DoS general}]
\begin{eqnarray}
		\rho_{\hat{\mathcal{S}}}=   \int\limits_{\mathds{T}^2 }  \frac{\rm d^2 k}{(2\pi)^2}\Tr \underline{S}\Big( \sum_i \Omega_{r_i k_i} + \epsilon^{ij}\epsilon^{lm}\Omega_{k_i r_l}\Omega_{k_j r_m}\Big.\notag\\
		\hspace{-10pt}\Big.+\Omega_{k_x k_y} B_{xy}\Big)\,.\hspace{10pt}
	\label{eq:2dDos}
\end{eqnarray}
When considering the charge degree of freedom, the last term corresponds to the Streda formula~\cite{streda1982theory}, i.e., $\partial_{B_{xy}}\rho_{\text{charge}}=c_1$, that relates the change of the density of states induced by an applied magnetic field to the \Fst Chern number in momentum space.
Hereafter, we assume vanishing magnetic fields for simplicity.
The remaining terms in Eq.~\eqref{eq:2dDos} are proportional to the momentum-position curvatures and become nontrivial when deformation fields are applied.

Calculating the total spinor-charge in an arbitrary region $\mathcal{C}$ in position space we find 
\begin{eqnarray}	\displaystyle\displaystyle	q_{\hS}  \ds= \int\limits_{\mathcal{C}}   \rm d^2 r 	\rho_{\hat{\mathcal{S}}}\displaystyle=\sum_i \Phi^{i}_1+\Phi_2\,.
		\label{eq:2dacc}
\end{eqnarray}
The first term corresponds to geometrical contributions from a set of \Fst spinor-Chern fluxes $\Phi^{i}_1$, defined over the respective $(r_i,k_i)$-sub-manifold [cf.~Eq.\eqref{def:flux1}],
\begin{eqnarray}
	\Phi^{i}_1  =\frac{1}{(2\pi)^2	}\int\limits_{\mathcal{C}, \mathds{T}^2} \rm d^2 r	\rm d^2 k \Tr \underline{S}\Omega_{r_i k_i}\,.
\end{eqnarray}
In similitude to the one-dimensional case [cf.~Eq.~\eqref{eq:1d spin accum}], these terms give rise to spinor-charge accumulation with co-dimensions 1 and can be generally quantized by global symmetry constraints.
An interesting manifestation of the \Fst spinor-Chern number in two dimensions are the helical edge state that appear at the boundaries of the insulator~\cite{liu2019helical,fang2019new}.
% , enabling a new quantum device made from a topological crystalline insulator nanorod, a “helical nanorod,” which has a quantized longitudinal conductance

In addition to the \Fst spinor-Chern fluxes, at second-order in the inhomogeneities we find a new geometrical contribution proportional to the \Snd spinor-Chern flux $\Phi_2$ defined in the entire momentum-position space $ (\bm k ,\bm r)$ [cf.~Eq.\eqref{def:flux2}],
\begin{eqnarray}
	\Phi_2 =\frac{1}{(2\pi)^2	}\int\limits_{\mathcal{C}, \mathds{T}^2}\rm d^2 r	\rm d^2 k \,\epsilon^{ij} \Tr  \underline{S}\Omega_{k_i r_x}\Omega_{k_j r_y}\,.
	\label{eq:2d flux}
\end{eqnarray}
Importantly, this quantity is intrinsically four-dimensional and vanishes for manifolds with dimensions three or less.
Since the integration region $\mathcal{C}$ is an open domain in the $(\bm k,\bm r)$ space, $\Phi_2$ is generally not expected to be quantized.
However, as we show below, under symmetry constraints it can become quantized and fractional.

The nontrivial effects of the \Snd spinor-Chern flux defined in Eq.~\eqref{eq:2d flux} manifest in the 2D model Hamiltonian of Eq.~\eqref{eq:2DHam PS} when the external parameters $\bm \phi$ depend on space.
Specifically, we assume $\phi_x$ ($\phi_y$) is only a function of $r_x$ ($r_y$) and takes continuous values between $0$ and $\pi$.
The integration domain $\mathcal{C}$ is assumed to cover the intersection of the two domain walls, see Fig~\ref{fig:fig2dpoles}(b).
In this case, the induced spinor-charge density has vanishing contributions from the \Fst spinor-Chern flux (both for charge and physical spin), as well as from the \Snd Chern flux associated to the charge degree.

In contrast, the \Snd spin-Chern flux associated to $\hat{\mathcal{S}}=\sigma_z\otimes\mathds{1}\otimes\mathds{1}$, is nonzero and has a closed analytic form
\begin{eqnarray}
\ds\frac{3}{8\pi^2}\int\limits_{ \mathds{T}^2\times \left[0,\pi\right]^2}\mathrm{d}^4 \mathbf{{\xi}}
	\hat{\mathbf{\textbf{d}}}\cdot (\partial_{\mathrm{k}_x}\hat{\textbf{\text{d}}} \times \partial_{\mathrm{k}_y}\hat{\mathbf{\textbf{d}}}\times \partial_{\phi_x}\hat{\mathbf{\textbf{d}}} \times \partial_{\phi_y}\hat{\mathbf{\textbf{d}}})\,,\notag
\end{eqnarray}
where $ \mathds{T}^2\times \left[0,\pi\right]^2$ is the integration domain in the four-dimensional parameter-space $\bm \xi = ( \bm k, \bm \phi)$. %~\cite{teo2010topological}
This expression is one of the main results of this paper: it connects an abstract geometrical property of phase-space -- the \Snd spin-Chern flux-- to a physical observable -- the spin accumulation.
Importantly, when chiral symmetry is restored, i.e., when $h_0 \to 0$, the \Snd spin-Chern flux and, hence, the accumulated spin, cf.~Eq.~\eqref{eq:2dacc}, becomes fractional and equal to
\begin{eqnarray}
	|q_\text{spin}| = \frac{1}{2}\,.
	\label{eq:2dSpinQuant}
\end{eqnarray}
The above equation is the extension of the fractional Berry flux attached to a 2D Dirac cone, cf.~Eq.~\eqref{eq:1stspinflux}, to the \Snd spin-Chern flux on the 4D Dirac-like cone supported in the $(\bm k , \bm \phi)$ parameter space, cf. Fig.~\ref{fig:fig2d}(c).

Comparing with the classical expectation of the multipole description of materials, the calculated spinor-charge density of Eq.~\eqref{eq:2dDos} must be created by the spatial gradient of the spinor-multipole moment densities, i.e.,
\begin{eqnarray}
	\rho_{\hat{\mathcal{S}}}=   - \nabla_{r} \cdot\bm{\mathcal{P}}^{\hat{\mathcal{S}}} + \frac{1}{2}\partial_i \partial_j \mathcal{Q}^{\hat{\mathcal{S}}}_{ij}(\mathbf{r})\,,
	\label{eq:2dClassicalPoles}
\end{eqnarray}
where repeating indices are summed, and $\nabla_{r}$ is a vector of spatial derivatives.
The multipole expansion of spinor-charge density allows for a geometrical interpretation in terms of the \Fst and \Snd spinor-Chern fluxes, namely,
\begin{eqnarray}
		\sum_i \Phi^{i}_1+	\Phi_2 \overset{!}{=}\int\limits_\mathcal{C} \rm  d^2r   \left(-\nabla_{r} \cdot\bm{\mathcal{P}}^{\hat{\mathcal{S}}} + \frac{1}{2} \partial_i \partial_j \mathcal{Q}^{\hat{\mathcal{S}}}_{ij}\right)\,.
	\label{eq:2dDoSfluxpoles}
\end{eqnarray}
The connection between spinor-Chern fluxes defined over the phase-space of the system and the accumulation of spinor-charge is a key outcome of this paper. 
Even though in general the definitions of multipole moments lack a geometrical expression, the semiclassical formalism provides a well-defined way to connect integrated differences of the multipoles to the geometrical properties of the system's phase-space. %, thus, eliminating the gauge ambiguity in their definition.

\subsection{In three dimensions}

In this Section, we generalize the concepts developed thus far to three-dimensional insulators.
Specifically, we calculate the response of a 3D material under spatio-temporal modulations and general external electromagnetic fields using a semiclassical approach valid up to third-order in perturbation theory.
We find that, alongside the \Fst and \Snd sub-spinor-Chern number responses encountered in Secs.~\ref{sec:1D} and \ref{sec:2D}, the spinor-current has a unique \Trd spinor-Chern number response associated to a topological index defined in the entire six-dimensional phase-space.
Similarly, the spinor-charge accumulation is shown to have contributions from a \Fst, \Snd and \Trd spinor-Chern fluxes that are associated to co-dimension 1, 2, and, 3 states, respectively.
Under symmetry constraints, we show that the \Trd spinor-Chern flux can become quantized, leading to a fractional spinor-charge accumulation localised at the corners of the three dimensional material.
Finally, we relate the spinor-Chern numbers and fluxes to the spatio-temporal modulations of the spinor-multipole moments. 

\begin{figure}
	\centering
	\includegraphics[width=1\linewidth]{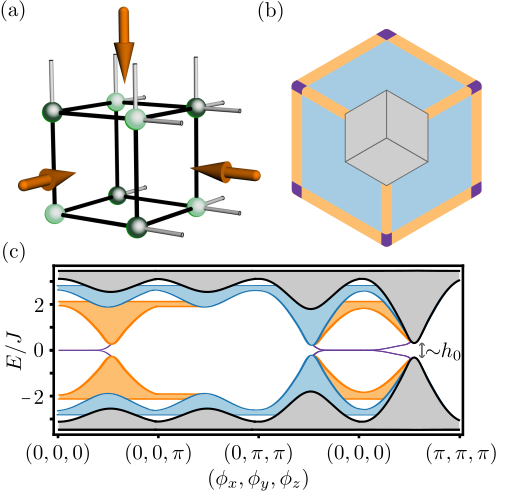}
	\caption{The model Hamiltonian in three dimensions. 
		(a) The 3D tight-binding lattice described by the Hamiltonian of Eq.~\eqref{eq:3DHam}.
		For simplicity, we only show hopping between equivalent spins.
		Light/dark green sites denote an onsite potential with positive/negative sign.
		Black (gray) lines denote hopping with amplitude $J_\mu ^+$ ($J_\mu ^-$).
		Each face of the cubic lattice is threaded by a strong magnetic field of $\pi$-flux quanta per plaquette.
		(b) The eigenstates of the open boundary spectrum are generally split into bulk (gray), surface (blue), hinge (orange), and corner (purple) states.
		(c) The open boundary spectrum of the Hamiltonian showing bulk (gray), surface (blue), edge (orange), and corner (purple) states.
		For the simulations we used $J_\mu /10=\Delta J^0 _\mu  = 0.1$, $h_0 =0.25$, and $|\bm \lambda^\mu|=0$.
		}
	\label{fig:fig3d}
\end{figure}

These concepts are illustrated in the spin responses of a concrete tight-binding model of spinful electrons on a cubic lattice.
Similar to the lower dimensional analogues discussed in Sections~\ref{sec:1D} and \ref{sec:2D}, the key ingredients in the Hamiltonian are the nearest-neighbour interaction, the dimerization of the hopping amplitudes in the three directions, the staggered on-site potential, and the spin-orbit coupling.
%Hence, the Hamiltonian in position space is given by $ H = H_\text{\text{KE}} + H_\text{d} + H_\text{h} +  H_\text{SO}$, where the appropriate generalization to three dimensions is done, see Appendix for more details.
%Here, we summarize the key ingredients.
%First, is the nearest neighbour interaction that determines the high-energy kinetic properties.
%Next, is the dimerization of the hopping amplitudes in the three directions and the staggered on-site potential that induce a gap in the spectrum.
%Lastly, is the spin-orbit coupling that hybridizes the states and lifts their degeneracy.
To keep the description less cumbersome, we directly use the momentum-space Hamiltonian [for an illustation of the real space crystal, see Fig.~\ref{fig:fig3d}(a)]
\begin{eqnarray}
	H(\bm k,\bm\phi) = \bm{d}\cdot \bm{\gamma} + \bm{D}\cdot\bm{\Gamma}\,,
	\label{eq:3DHam}
\end{eqnarray}
where $\bm d$ and $\bm D$ represent a 7- and 18-component vector, respectively.
As these expressions are straight forward generalizations of Eq.~\eqref{eq:2DHam PS}, here, we only show the corresponding matrices
\begin{eqnarray}
	\begin{array}{rl}
		\bm{\gamma} = &\{\mathds{1}\otimes \sigma_x\otimes \mathds{1}\otimes\sigma_x,
		\mathds{1}\otimes \sigma_x\otimes \mathds{1}\otimes\sigma_y,\\
		&\,\,\, \mathds{1}\otimes \sigma_x\otimes \mathds{1}\otimes\sigma_z,\mathds{1} \otimes \sigma_y\otimes \sigma_z\otimes\mathds{1},\\
		&\,\,\, \mathds{1}\otimes \sigma_y\otimes \sigma_y\otimes\mathds{1}, \mathds{1} \otimes \sigma_y\otimes \sigma_x\otimes\mathds{1},\\
		&\,\,\, \sigma_z\otimes \sigma_z\otimes \mathds{1}\otimes\mathds{1} \}\,,
	\end{array}
\end{eqnarray}
that define the kinetic and potential energy, as well as the 18 matrices representing the spin-orbit interaction 
\begin{eqnarray}
	\begin{array}{rl}
		\bm{\Gamma} = &\{\sigma_x\otimes \sigma_x\otimes \mathds{1}\otimes\sigma_x,
		\sigma_x\otimes \sigma_x\otimes \mathds{1}\otimes\sigma_y,\\ &\,\,\,\sigma_y\otimes \sigma_x\otimes \mathds{1}\otimes\sigma_x,
		\sigma_y\otimes \sigma_x\otimes \mathds{1}\otimes\sigma_y,\\ &\,\,\,\sigma_z\otimes \sigma_x\otimes \mathds{1}\otimes\sigma_x,
		\sigma_z\otimes \sigma_x\otimes \mathds{1}\otimes\sigma_y,\\
		&\,\,\, \sigma_x\otimes \sigma_x\otimes \mathds{1}\otimes\sigma_z,\sigma_x\otimes \sigma_y\otimes \sigma_z\otimes\mathds{1},\\
		&\,\,\, \sigma_y\otimes \sigma_x\otimes \mathds{1}\otimes\sigma_z,\sigma_y\otimes \sigma_y\otimes \sigma_z\otimes\mathds{1},\\
		&\,\,\, \sigma_z\otimes \sigma_x\otimes \mathds{1}\otimes\sigma_z,\sigma_z\otimes \sigma_y\otimes \sigma_z\otimes\mathds{1},\\
		&\,\,\, \sigma_x\otimes \sigma_y\otimes \sigma_y\otimes\mathds{1}, \sigma_x\otimes \sigma_y\otimes \sigma_x\otimes\mathds{1},\\
		&\,\,\, \sigma_y\otimes \sigma_y\otimes \sigma_y\otimes\mathds{1}, \sigma_y\otimes \sigma_y\otimes \sigma_x\otimes\mathds{1},\\
		&\,\,\, \sigma_z\otimes \sigma_y\otimes \sigma_y\otimes\mathds{1}, \sigma_z\otimes \sigma_y\otimes \sigma_x\otimes\mathds{1} \}\,.
	\end{array}
\end{eqnarray}

We further assume that $H(\bm k, \bm \phi)$ describes a material where the dimerization parameters and on-site potential can depend both on space and time.
Formally, this is implemented by a set of external parameters $\bm \phi = (\phi_x ,\phi_y ,\phi_z)$, where
\begin{eqnarray}
	\Delta J_\mu = \Delta J_\mu ^0 \cos(\phi_\mu)\text{   and   }h  = h_0\prod_\mu\sin(\phi_\mu)\,,
\end{eqnarray}
with $\Delta J_\mu ^0$ and $h_0$ constants.
The spatio-temporal modulations of $\bm \phi$ are assumed to be weak enough such that the wavepacket's equations of motion are well-captured by third-order corrections, cf. Sec.~\ref{sec:Theory}.% and Appendix.~\ref{apx:semi}

The bulk spectrum of the Hamiltonian is composed by 16 bands which are split into two sets of positive and negative energies.
Each set can be further split into two quadruplets which are mixed depending on the spin-orbit interaction.
When $|\bm \lambda^\mu| = 0$, all positive (similarly, the negative) energy states become degenerate, while this is lifted to isolated regions in the Brillouin zone when $|\bm \lambda^\mu|> 0$, for any index $\mu$.
The material at half filling is conducting only when $\Delta J_\mu = h = 0$, for all $\mu$ indices.

The open boundary spectrum of the Hamiltonian has four distinct sets of states classified depending on their co-dimensionality, see Fig.~\ref{fig:fig3d}(b) and (c).
First, are the co-dimension 0 states, i.e., bulk modes, which correspond to fully delocalised wavefunctions that have nonzero probability on lattice points deep within the bulk of the material; these correspond to the solutions of the Bloch Hamiltonian of Eq.~\eqref{eq:3DHam}.
Next, are the co-dimension 1 states that are localized in one of the coordinates but extended in the remaining two; these states are found on the surfaces of the 3D material.
Then, are the co-dimension 2 states, which are localised in two dimensions but extended in the third, e.g., spin-helix hinge states.
Lastly, are co-dimension 3 states associated to fully localized states; such states appear on the corners of the material.
As a function of the external parameters $\bm \phi$, the boundary states disperse, merge into other bands and re-emerge according to the lattice parameters; however, only the co-dimension 3 states cross the gap.

To conclude the description of the model, we note that the Hamiltonian has chiral symmetry $\chi = \mathds{1}\otimes\sigma_z\otimes\mathds{1}\otimes\mathds{1}$ and T.R. symmetry $T=i\mathds{K}\sigma_y\otimes\mathds{1}\otimes\mathds{1}\otimes\mathds{1}$ only when $h=0$.
On the other hand, the parameter space has a global T.R. symmetry $T$ for any value of the $\Delta J_\mu$, $\bm\lambda^\mu$ and $h$.
% The parameter space has charge conjugation symmetry $C=i\mathds{K}\sigma_y\otimes\sigma_z\otimes\mathds{1}\otimes\mathds{1}$ for any value of the $\Delta J_\mu$, $\bm\lambda^\mu$ and $h$ that ascribes a $\mathds{Z}$ topological index~\cite{altland1997nonstandard}.

% a unitary T.R. symmetry $\Theta'=i\mathds{K}\sigma_x\otimes\mathds{1}\otimes\mathds{1}\otimes\mathds{1}$ for any value of the $\Delta J_\mu$, $\bm\lambda^\mu$ and $h$ that ascribes a $\mathds{Z}_2$ topological index~\cite{altland1997nonstandard}.

\subsubsection{Transport}

Having seen the relation between spinor-charge transport in weakly perturbed materials and spinor-Chern numbers in dimensions one and two, cf.~Secs.~\ref{sec:1Dtrans} and \ref{sec:2Dtrans}, we are now in a position to discuss topological transport in three dimensions and how it relates to the spinor-Chern numbers. 
Extending previous results for the charge degree~\cite{petrides2018six}, we show how the adiabatic evolution of the Hamiltonian induces a third-order correction proportional to the \Trd spinor-Chern number.
Alongside this unique response, we find a set of lower-dimensional indices, the \Fst and \Snd sub-spinor-Chern numbers, that appear as first and second order corrections to the spinor-charge current.
Additionally, by including nonzero external electromagnetic fields we derive Axion responses in the spinor degrees of freedom.
Finally, we show how our results manifest in the modulations of the spinor-multipole moments.

Using the semiclassical theory developed in Section~\ref{sec:Theory}, a three-dimensional insulator in the absence of external electromagnetic fields is characterized by the spinor-current
\begin{eqnarray}
	\begin{array}{rl}
	\ds	j^{i}_{\hS} =  \int\limits_{\mathds{T}^3 } \frac{\rm d^3 k}{(2\pi)^3}\Tr\uS\left(\right.&\ds\Omega_{t k_i} +\epsilon^{ijk} \epsilon^{klm} \Omega_{t k_l}\Omega_{r_j k_m}  \\
		&\ds\left.+\epsilon^{ijk} \epsilon^{lmn} \Omega_{t k_l}\Omega_{r_j k_m}\Omega_{r_k k_n} \right)\,,\,\,\,\,\,
	\end{array}
	\label{eq:3dresult}
\end{eqnarray}
where the integration domain is over a $\mathds{T}^3$ torus representing the 3D Brillouin zone, and Latin indices run over three directions $i \in\{x,y,z\}$.  
Depending on the particular Hamiltonian, the derived spinor-current of Eq.~\eqref{eq:3dresult} includes a variety of phenomena.
The first term is equivalent to Eq.~\eqref{eq:1dresult} and results in spinor-charge transport proportional to a \Fst sub-spinor-Chern number defined in the $(t,k_i)$ manifold.
The next term is a double product of curvatures and gives rise to 2D topological spinor pumps with a \Snd sub-spinor-Chern number response [cf. Eq.~\eqref{eq:2dresult}].
Finally, the last term is a unique three-dimensional response given by a triple product of curvatures in the entire $(\bm k, \bm r ,t )$ phase space.

Next, when nonzero external electromagnetic fields are imposed, we derive two additional corrections to Eq.~\eqref{eq:3dresult}.
At first order we obtain
\begin{eqnarray}
	\int\limits_{\mathds{T}^3 } \frac{\rm d^3 k}{(2\pi)^3}\frac{1}{2}\epsilon^{ijl}\epsilon^{lmn}\Tr\uS\ds\Omega_{k_m k_n}E_j\,,
\end{eqnarray}
corresponding to the previously encountered spinor-Hall (or spin-Hall, depending on the chosen degree of freedom) response, cf.~Eq.~\eqref{eq:2dresult}, that relates the application of an electric field to a perpendicular spinor-current with proportionality constant the \Fst spinor-Chern number in momentum space.
At second order, the corrections are given by
%\begin{eqnarray}
 %   \begin{array}{rl}
 %   \int\limits_{\mathds{T}^3 } \frac{\rm d^3 k}{(2\pi)^3}\Tr\uS\left(\right.&\ds\epsilon^{ijk} \epsilon^{lmn} E_{j}\Omega_{k_l k_m}\Omega_{r_k k_n}\\
%		&\ds\left.+\epsilon^{ijk} \epsilon^{lmn} \Omega_{t k_l}\Omega_{k_m k_n}B_{jk} \right)\,,\,\,\,\,\,
%	\end{array}
%\end{eqnarray}
\begin{eqnarray}
    \begin{array}{rl}
     \epsilon^{ilm}E_{l}\partial_{r_m}\theta+\epsilon^{ilm} B_{lm}\dot{\theta} \,,\,\,\,\,\,
	\end{array}
\end{eqnarray}
where
\begin{eqnarray}
    \theta = \int\limits_{\mathds{T}^3 } \frac{\rm d^3 k}{(2\pi)^3}\epsilon^{lmn}\Tr\uS  \Omega_{k_l k_m}A_{k_n} \,,
    \label{eq:Axion}
\end{eqnarray}
is dubbed the ``spinor-Axion index".
Similar to the usual charge responses due to a nontrivial Axion field~\cite{QiZhang}, the simultaneous application of an external electric (magnetic) field and the spatial (temporal) modulations of the Hamiltonian induce a nontrivial spinor-current that depends on a topological property of the combined momentum-position-time coordinates.
As these results directly generalize previous findings that only considered the charge degree~\cite{QiZhang}, for simplicity, hereafter we assume no external electromagnetic fields such that spinor-Axion responses vanish.

Integrating the spinor-current in the $i$-th direction over a full pump cycle we obtain the spinor-charge transport 
\begin{eqnarray}
	\Delta q_{\hS}^i  =  \int\limits_{T,V }   {\rm d}t{\rm d}^3  r j^{i}_{\hS} = c^{i}_1+ c^{i}_2+ c^i_3\,,
	\label{eq:3d pumped charge}
\end{eqnarray}
where the integral runs over a period $T$ and the volume of the unit cell $V$ (both set to unity for simplicity).
The first two contributions are already derived in the context of one- and two-dimensional systems [cf. Eqs.~\eqref{eq:pumped charge} and~\eqref{eq:2d pumped charge}] and in three dimensions are proportional to the \Fst and \Snd sub-spinor-Chern numbers, namely,
\begin{eqnarray}
	c^{i}_1 = \frac{1}{(2\pi)^3} \int\limits_{T,V ,\mathds{T}^3 }   {\rm d}t{\rm d}^3r {\rm d}^3 k \Tr\uS\Omega_{t k_i}\notag\,,
\end{eqnarray}
and
\begin{eqnarray}
	c^i_2 = \frac{1}{(2\pi)^3} \int\limits_{T,V ,\mathds{T}^3 }   {\rm d}t{\rm d}	^3r {\rm d}^3 k \,\epsilon^{ijk} \epsilon^{klm} \Tr\uS\Omega_{t k_l}\Omega_{r_j k_m}\notag\,. 
\end{eqnarray}
In addition to these responses, at third order in perturbation theory, a \Trd spinor-Chern number response manifests, defined as
\begin{eqnarray}
	c^i_3 = \frac{1}{(2\pi)^3} \int\limits_{T,V ,\mathds{T}^3 }   \rm dt\rm d^3r \rm d^3 k \epsilon^{ijk} \epsilon^{lmn}\Tr\uS \Omega_{t k_l}\Omega_{r_j k_m}\Omega_{r_k k_n}\notag\,.
\end{eqnarray}
The above expression is indeed the \Trd spinor-Chern number characterizing the $(\bm k , \bm r , t)$ phase-space since contributions from electromagnetic fields are assumed to be vanishing.

The derived expression of spinor-charge transport, Eq.~\eqref{eq:3d pumped charge}, can be decomposed as the temporal gradient of the spinor-dipole moment density, the second derivative of the spinor-quadrupole moment and the third derivative of the spinor-octapole moment
\begin{eqnarray}
j^{i}_{\hS}  =   - \partial_{t} {\mathcal{P}^{\hS}_i} +\partial_{t} \partial_{r_j} \mathcal{Q}^{\hS}_{ij}(\mathbf{r}) + \frac{1}{2}\partial_{t} \partial_{r_j}\partial_{r_l} \mathcal{O}^{\hS}_{ijl}(\mathbf{r})\,,
	\label{eq:multi curent 3}
\end{eqnarray}
where $\bm{\mathcal{P}}^{\hS}$, $\mathcal{Q}^{\hS}_{ij}$, and, $\mathcal{O}^{\hS}_{ijl}$ are the spinor analogues of the electric dipole, quadrupole, and octapole moment densities.
This decomposition can be used as an alternative definition of the spinor-multipole moments
\begin{align}
	\begin{array}{rl}		
		c^{i}_1+ c^i_2+ c^i_3
		\overset{!}{=}\ds \int\limits_{T,V  }   \rm dt\rm d^3r   &\ds\left( -\partial_{t} {\mathcal{P}^{\hS}_i} +\partial_{t} \partial_{r_j} \mathcal{Q}^{\hS}_{ij}\right.\\
		&\ds\,\,\,\left.+ \frac{1}{2}\partial_{t} \partial_{r_j}\partial_{r_l} \mathcal{O}^{\hS}_{ijl}\right)\,,
	\end{array}	\label{eq:geomul3}
\end{align}
i.e., integrated differences of spinor-multipole moments are determined by the spinor topological properties of phase-space.

Focusing on the particular three-dimensional model of Eq.~\eqref{eq:3DHam}, the topological aspects of phase-space become nonvanishing when $\bm \phi$ is a function of both space and time.
Specifically, we assume $\phi_x (t)$ is a function of only time and takes values in the interval $\left[0,2\pi\right]$, while $\phi_y$ ($\phi_z$) is a function of only $x$ ($y$) and is smoothly varied between $0$ and $2\pi$.
In this case, the \Fst and \Snd sub-spinor-Chern numbers of the system vanish for both the charge and physical spin degree, i.e., for $\hS = \mathds{1}\otimes\mathds{1}\otimes\mathds{1}\otimes\mathds{1}$ or $\hS = \sigma_z\otimes\mathds{1}\otimes\mathds{1}\otimes\mathds{1}$ respectively.
% Equivalently, the \Trd Chern number associated to the charge degree is also zero.
The only nonvanishing contribution comes from the \Trd spinor-Chern number and is given by 
\begin{eqnarray}
	\begin{array}{rl}		
			\Delta q_{\hS}^i   &\ds=\frac{5}{4\pi^3}\int_{\mathds{T}^3\times\left[0,2\pi\right]^3}\mathrm{d}^3 {{k}}\mathrm{d}^3 \phi\,\delta_{i,z}\Tr \uS \sigma_z\otimes\mathds{1}\otimes\mathds{1}\otimes\mathds{1}\\\\
			&\,\,\,\hat{\mathbf{\textbf{d}}}\cdot (\partial_{\mathrm{k}_x}\hat{\textbf{\text{d}}} \times \partial_{\mathrm{k}_y}\hat{\mathbf{\textbf{d}}}\times \partial_{\mathrm{k}_z}\hat{\mathbf{\textbf{d}}}\times \partial_{\phi_x}\hat{\mathbf{\textbf{d}}} \times \partial_{\phi_y}\hat{\mathbf{\textbf{d}}}\times \partial_{\phi_z}\hat{\mathbf{\textbf{d}}}) \,,
	\end{array}\notag
\end{eqnarray}
where $\hat{\bm d}=\bm d/|\bm d|$ and $\delta_{i,j}$ is the Dirac delta function.
As in the 2D case, the later stems from the Jacobian of the transformation between the position-time manifold and the $\bm\phi$ parameter space.
In the case where $\hat{S}$ corresponds to the charge degree of freedom, the above expression vanishes because of the trace.
On the other hand, when $\hat{S} =  \sigma_z\otimes\mathds{1}\otimes\mathds{1}\otimes\mathds{1}$, i.e., for physical spin, the contribution from the \Trd spin-Chern number becomes quantized and equal to $4$.

\begin{figure}
	\centering
	\includegraphics[width=1\linewidth]{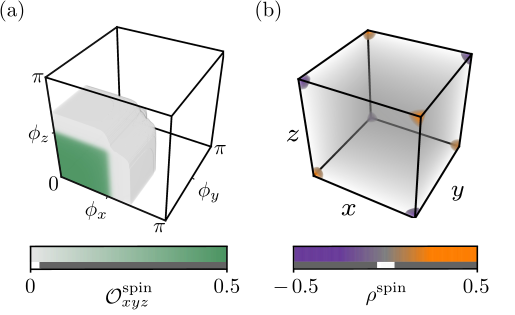}
	\caption{Spin-octapole moment. 
		(a) The spin-octapole moment, cf. Appendix~\ref{apx:poles}, as a function of the external parameters $\bm \phi$ takes values between $0$ (transparent) and $1/2$ (green).
		The transition between the two values (gray region) is controlled by the chiral breaking mass $h_0$, i.e., the smaller it is, the sharper the transition.
		(b) The electronic ground state of the open boundary crystal at $ \phi = (0,0,0)$ exhibits spin localization at the corners, as expected from a non vanishing spin-octapole moment.
		We note that the analogous expressions for the charge degree of freedom vanish independent of $ \phi$.
		For the simulations we have used $\Delta t^0 _\mu / t = 0.1$, $h/t =0.01$, and $|\bm D|=0$.
		}
	\label{fig:fig3dpoles}
\end{figure}

The nonzero transport of physical spin is reflected in the modulations of the spin-multipole moments, shown in Fig.~\ref{fig:fig3dpoles}(a).
As a function of the external parameters, the spin-octapole moment takes continuous values and ``winds" around singular points in the $\bm \phi $ parameter space.
This leads to a nonzero third gradient and to a nontrivial contribution to spin-current, as described by Eq.~\eqref{eq:multi curent 3}.
We note that all other (both electric and spin) multipole moments vanish due to global and local symmetry constraints.

\subsubsection{Accumulation}

We, now, derive the spinor-charge density using a semiclassical approach valid up to third order in perturbation theory.
At first order, we find a set of \Fst spinor-Chern fluxes that are related to co-dimension 1 states, as already discussed in Sec.~\ref{sec:1Dacc}. 
In addition, we obtain the generalization of the Streda formula to three dimensions and the relation of spinor-charge density to the \Fst spinor-Chern number in momentum space. 
Second order corrections are given by a set of \Snd spinor-Chern fluxes that give rise to co-dimension 2 states, cf. Section~\ref{sec:2Dacc}.
Next, are the spinor-Axion field responses that depend on the applied electromagnetic fields, as well as on the deformation fields.
Such corrections give rise to the generalized magneto-electric effect~\cite{QiZhang} which relates spinor-charge localization to the application of a parallel magnetic field.
Lastly, the \Trd spinor-Chern flux appears as a unique third order correction and is related to states with co-dimension 3.
Extending the electric multipole description to spinor degrees of freedom and including effects up to the spinor-octapole moment, we establish a fundamental relation between boundary states, spinor-multipole moments, and, the geometrical properties of phase-space.

The spinor-charge density of a generic 3D insulator under arbitrary perturbing fields is calculated using Equation~\eqref{eq:density}.
As this expression contains numerous terms we first show the corrections that depend on the electromagnetic fields and then focus on pure deformation fields.
At first order, we obtain the 3D analogue of the Streda formula, namely,
\begin{eqnarray}
   	\int\limits_{\mathds{T}^3 } \frac{\rm d^3 k}{(2\pi)^3}\frac{1}{2} \Tr\uS\Omega_{k_i k_j}B_{ij}\,,
\end{eqnarray}
that relates the spinor-change density to the applied magnetic field and to the \Fst sub-spinor-Chern number in momentum space, cf.~Sec.~\ref{sec:2Dacc}.
Next, we derive the spinor-Axion response
\begin{eqnarray}
\frac{1}{2}   \epsilon^{ijk} \partial_{r_i}\theta B_{jk}\,,
\end{eqnarray}
that gives rise to a nonzero spinor-charge density depending on the gradient of the spinor-Axion field $\theta$, defined in Eq.~\eqref{eq:Axion}.

The remaining corrections due to deformation fields are given by
\begin{eqnarray}
\ds
    \rho_{\hS}=  \int\limits_{\mathds{T}^3 } \frac{\rm d^3 k}{(2\pi)^3}\Tr\uS\left(\ds\sum\limits_i\Omega_{r_i k_i} +\sum\limits_{i,j}\epsilon^{ikl} \epsilon^{jmn} \Omega_{r_k k_m}\Omega_{r_l k_n} \right. \notag\\
		 +\epsilon^{ijk} \epsilon^{lmn} \Omega_{r_i k_l}\Omega_{r_j k_m}\Omega_{r_k k_n}\Bigg)\,.
\notag\\
    \label{eq:3Ddensity}
\end{eqnarray}
The first two terms have already been encountered in Secs.~\ref{sec:1Dacc} and~\ref{sec:1Dacc}, albeit from a lower dimensional perspective; in three dimensions, these terms can lead to helical surface and hinge states, respectively~\cite{song2017d,geier2018second}.
On the other hand, the last term in Eq.~\eqref{eq:3Ddensity} is an intrinsically three dimensional response as it depends on the full six-dimensional phase-space manifold.

The accumulation of spinor-charge in an arbitrary region $\mathcal{C}$ in position space is, hence, given by
\begin{eqnarray}	\displaystyle\displaystyle	q_{\hS}  \ds= \int\limits_{\mathcal{C}}   \rm d^2 r 	\rho_{\hat{\mathcal{S}}}\displaystyle=\sum_i \Phi^{i}_1+\sum\limits_{i,j}\Phi^{ij}_2+\Phi_3 \,.
		\label{eq:3Dacc}
\end{eqnarray}
The first term defines a set of \Fst spinor-Chern fluxes $\Phi^{i}_1$ in the $(r_i,k_i)$-sub-manifold [cf.~Eq.\eqref{def:flux1}],
\begin{eqnarray}
	\Phi^{i}_1  =\frac{1}{(2\pi)^3	}\int\limits_{\mathcal{C}, \mathds{T}^3} \rm d^3 r	\rm d^3 k \Tr \underline{S}\Omega_{r_i k_i}\,.
\end{eqnarray}
and induces a spinor-charge accumulation with co-dimensions 1.
Next, is the \Snd spinor-Chern flux $\Phi^{ij}_2$ defined in a four-dimensional sub-manifold of phase-space [cf.~Eq.\eqref{def:flux2}],
\begin{eqnarray}
	\Phi^{ij}_2 =\frac{1}{(2\pi)^3	}\int\limits_{\mathcal{C}, \mathds{T}^3}\rm d^3 r	\rm d^3 k \epsilon^{ikl} \epsilon^{jmn}\Tr\uS \Omega_{r_k k_m}\Omega_{r_l k_n}\,.
	\label{eq:3d flux}
\end{eqnarray}
Similar to the two-dimensional case, cf.~Eq~\eqref{eq:2dacc}, the \Snd spinor-Chern flux appears as a second-order correction and is related to states localised in two coordinates but extended in the remaining.
Lastly, at third order we obtain a unique three-dimensional response related to the \Trd spinor-Chern flux
\begin{eqnarray}
    \Phi_3 =\int\limits_{\mathcal{C}, \mathds{T}^3}\frac{\rm d^3 r	\rm d^3 k}{(2\pi)^3	}\epsilon^{ijk} \epsilon^{lmn}  \Tr\uS\Omega_{r_i k_l}\Omega_{r_j k_m}\Omega_{r_k k_n}\,.
\end{eqnarray}
As the integration region $\mathcal{C}$ does not necessarily cover the entire parameter space, the corrections $\Phi_1$, $\Phi_2$ and $\Phi_3$ are quantized only when symmetry constraints are imposed.

Focusing on our tight-binding model~\eqref{eq:3DHam}, the nontrivial geometrical properties of phase-space manifest in the physical spin accumulation when the parameters $\bm \phi$ are properly modulated in space.
Here, we assume that each $\phi_i$ is a function only of the associated position coordinate, i.e., $\partial_{r_i}\phi_j = \delta_{i,j}$, and takes values between $0$ and $\pi$ within a finite region.
Furthermore, we take $\mathcal{C}$ to be the support of the three-dimensional domain wall defined by the gradients of the external parameters $\bm\phi$.
Calculating the corrections associated to the charge degree of freedom, i.e., taking $\hS = \mathds{1}\otimes\mathds{1}\otimes\mathds{1}\otimes\mathds{1}\otimes\mathds{1}$, we find that all Chern fluxes vanish.
This is also the case for the \Fst and \Snd spin-Chern fluxes associated to physical spin, i.e., when $\hS = \sigma_z\otimes\mathds{1}\otimes\mathds{1}\otimes\mathds{1}\otimes\mathds{1}$.

The only surviving term is the \Trd spin-Chern flux, given by

\begin{eqnarray}
	\begin{array}{rl}		
			\Phi_3  &\ds=\frac{5}{4\pi^3}\int_{\mathds{T}^3\times \left[0,\pi\right]^3}\mathrm{d}^3 {{k}}\mathrm{d}^3 \phi\\\\
			&\,\,\,\hat{\mathbf{\textbf{d}}}\cdot (\partial_{\mathrm{k}_x}\hat{\textbf{\text{d}}} \times \partial_{\mathrm{k}_y}\hat{\mathbf{\textbf{d}}}\times \partial_{\mathrm{k}_z}\hat{\mathbf{\textbf{d}}}\times \partial_{\phi_x}\hat{\mathbf{\textbf{d}}} \times \partial_{\phi_y}\hat{\mathbf{\textbf{d}}}\times \partial_{\phi_z}\hat{\mathbf{\textbf{d}}}) \,,
	\end{array}\notag
\end{eqnarray}
where $\mathds{T}^3\times \left[0,\pi\right]^3$ is the integration domain in the six-dimensional parameter-space $\bm \xi = ( \bm k, \bm \phi)$.
The expression of the \Trd spin-Chern flux becomes quantised and equal to $1/2$ when chiral symmetry is imposed to the tight-binding model, i.e., when $h_0 \to 0$.
As a result, the accumulation of spin at the 0D boundary defined by the domain wall becomes
\begin{eqnarray}
	|q_{\text{spin}}| = \frac{1}{2}\,.
\end{eqnarray}

Similar to Eq.~\eqref{eq:2dDoSfluxpoles}, an alternative interpretation of the accumulated spin (and in general spinor-degrees of freedom) is obtained by the classical theory of multipole moments.
Within this description, the spinor-Chern fluxes are related to the modulations of, what we dub, spinor-multipole moments
\begin{align}
	\begin{array}{rl}		
		&\ds\sum_i \Phi^{i}_1+\sum\limits_{i,j}\Phi^{ij}_2+\Phi_3\\ 
		\overset{!}{=}&\ds \int\limits_{T,V  }   \rm dt\rm d^3r   \ds\left( -\partial_{t} {\mathcal{P}^{\hS}_i} + \frac{1}{2}\partial_{t} \partial_{r_j} \mathcal{Q}^{\hS}_{ij}\right.\ds\,\,\,\left.+ \frac{1}{6}\partial_{t} \partial_{r_j}\partial_{r_l} \mathcal{O}^{\hS}_{ijl}\right)\,.
	\end{array}
\end{align}
Indeed, the bulk spin-octapole moment of the tight-binding model acquires nonzero values depending of the external parameters $\bm\phi$, cf.~Fig.~\ref{fig:fig3dpoles}.
Around the high symmetry point $\bm \phi=(\pi/2,\pi/2,\pi/2)$, the third gradient of the spin-octapole moment diverges depending on the value of the chiral breaking mass $h_0$; in the limit where $h_0\sim 0$, its contribution to the spin density becomes quantized and equal to $1/2$, as predicted by Eq.~\eqref{eq:3Dacc}.

\section{Conclusions}

Designing realistic materials that can be easily controlled is of paramount importance when proposing experiments. 
Multiferroic materials provide a promising platform for controlling electronic properties with external electromagnetic and deformation fields~\cite{eerenstein2006multiferroic}.
% In recent years, the multiferroic properties of materials have shown capable of tremendous control over electron dynamics.
In particular, ferromagnetic compounds with alternating crystal axes, such as Cu-benzoate~\cite{affleck1999field} and $Yb_4As_3$~\cite{Kohgi01}, develop a staggered onsite potential when a perpendicular uniform magnetic field is applied and, as a consequence, the material becomes insulating.
This is due to the competition between Dzyaloshinskii-Moriya interaction and a nonzero gyromagnetic tensor.
The former can also give rise to an exchange interaction that depends on the applied electric field, as demonstrated in $MnI_2$ and oxides $ABO_2$ with $A =  Cu$, $Ag$, $Li$ or $Na$ and $B=Cr$ or $Fe$~\cite{xiang2011general,kimura2006inversion,seki2007impurity,seki2008spin}. 
More related to this paper, a highly efficient control of the antiferromagnetic order using a uniform magnetic field was demonstrated in two-dimensional latices of $SrIrO_3$ and $SrTiO_3$~\cite{hao2018giant}.

Coupling electronic properties to strain offers an alternative route towards inducing controlled dynamics.
Specifically, materials with piezo-electric, piezo-magnetic or flexo-electric properties develop nonzero electric and magnetic moments, such as polarisation and magnetisation, in response to strain~\cite{kogan1964piezoelectric,gopalan2007defect,zubko2013flexoelectric,hao2018giant,tagantsev1986piezoelectricity,yudin2013fundamentals}.
Furthermore, symmetry analysis revealed an interesting class that combines electric and magnetic properties to give the ``piezo-magnetoelectric effect''~\cite{grimmer1992piezomagnetoelectric,benveniste1995magnetoelectric}, i.e., the material develops a nonzero polarisation due to a parallel magnetic field and strain.
Enhanced piezo-magnetoelectric properties were also observed in ceramics, rare-earth iron alloys, polymer composites~\cite{van1974situ,nan2002three}, laminates~\cite{ryu2001magnetoelectric,cai2004large,srinivasan2001magnetoelectric}, and epitaxial multilayers~\cite{lee2000strain}.
These materials have seen an enormous use in applications, both for their fundamental interest as well as their practicality.

Our semiclassical treatment of electrons in insulating crystals establishes a natural description of topological aspects, as it mostly arises from wave interference.
As we have already shown, the nontrivial geometrical structure of phase-space is fundamentally connected to macroscopic responses, namely, transport and accumulation.
Such connection ultimately provides an alternative definition of the electric multipole moments in the form of integrated differences, thus, eliminating any gauge ambiguity that can arise from the boundary conditions.

Introducing new internal quantum degrees of freedom augments the semiclassical description and leads to new topological constructs. 
This works provides an exhaustive delineation of such a generalized semiclassical theory, while deriving a novel set of multipoles with internal structure -- the spinor-multipole moments. 
We believe that our work will inspire and guide novel solid state studies both in real materials and quantum engineered systems.
Of particular interest are the large-scale applications in quantum information technologies using qudits -- a multi-level alternative to the conventional 2-level qubit -- that are expected to provide an unprecedented storage capacity, processing power, secure
encryption, as well as, reducing circuit complexity, and increasing algorithm efficiency~\cite{moreno2018molecular,wang2020qudits}.

In this paper, we present a complete description of non-interacting electrons in weakly inhomogeneous, adiabatically driven insulators under external electromagnetic fields. 
We calculate the transport and accumulation of general spinor degrees of freedom using a semiclassical approach where we include corrections up to third-order in perturbation theory.
As such, we illustrate fundamental connections among geometry and physical observables that enable us to predict exotic states of matter.
The derived effects are studied in concrete tight-binding models where the aforementioned relations are calculated both analytical and numerically. 
Remarkably, our approach puts topological spinor pumps, the spinor-Hall effect, spinor-higher-order TIs, spinor-multipole moments, and spinor-Axion responses under the unifying umbrella of \textit{phase-space topology}.

\section{Acknowledgements}

We acknowledge useful discussions with Johannes Kellendonk. 
Work by I.P. is supported by the Early Postdoc mobility grant from the Swiss National Science Foundation (SNSF) under project ID P2EZP2\_199848.

\appendix
\section{Wavepacket construction}\label{apx:semi}

We start by formally expanding the Hamiltonian in distances $\delta \hat{\bs{r}} = \hat{\bs{r}}-\bs{r}$ around the center of mass position $\bs{r}$ as 
\begin{eqnarray}
	\hat{H} = \hat{H}_0 +\hat{H}' \,,
\end{eqnarray}
where $\hat{H}_0 =  \hat{H}(\bs r, \bs k , t)$ is the local Hamiltonian evaluated at the center of mass coordinates, and, $\hat{H}'$ are higher order corrections. 
For example, at third order this is given by (hereafter, we take $\hbar\!=\!e\!=\!1$)
\begin{eqnarray}
	\begin{array}{rl}
		\ds\hat{H}' =&\ \ds\frac{\partial \hat{H}_0}{\partial r_i} \delta \hat{\bs{r}}^{i}+ \frac{1}{2}\frac{\partial \hat{H}_0}{\partial r_i}\frac{\partial \hat{H}_0}{\partial r_j}  \{\delta \hat{\bs{r}}^i , \delta \hat{\bs{r}}^j\}\\
		&\ds+ \frac{1}{4}\frac{\partial \hat{H}_0}{\partial r_i}\frac{\partial \hat{H}_0}{\partial r_j}\frac{\partial \hat{H}_0}{\partial r_k}  \{\{\delta \hat{\bs{r}}^i , \delta \hat{\bs{r}}^j\},\delta \hat{\bs{r}}^k\}\,. 
		\label{eq:Ham exp}
	\end{array}
\end{eqnarray}
Regardless of their physical origins, the strength of these corrections eventually determines the choice of basis for the construction of the wavepacket.
Specifically, the size of the wavepacket must be much smaller than the characteristic length-scale defined by the $\hat{H}'$; both in the position and momentum space, see Fig.~\ref{fig:figsemi}. 
Constructing a wavepacket that is several orders of magnitude larger than the lattice constant, ensures a local basis of states with well-defined center-of-mass phase-space coordinates $(\bs r , \bs k )$, and where intermediate length scales are encoded perturbatively up to a sufficiently large order. 
Any other strong corrections are included intrinsically in the local Hamiltonian $\hat{H}_0$~\cite{Xiao2010, price2016measurement}.  

Concretely, the wavepacket is built directly from the $N$ eigenstates $\{\ket{\tilde{n}(\bs{k},\bs{r})}\}$ of a set of isolated energy bands of $\hat{H}$ up to a particular order~\cite{Niu1995,Chang1995,culcer2005coherent, Price2015,petrides2018six}, e.g., at second order in perturbation theory, the eigenstates are expanded as	
\begin{eqnarray}
	\ket{\tilde{n}} \approx \ket{n} +\ket{n'} + \ket{n''} + ... \,,
	\label{eq:pertbasis}
\end{eqnarray}
where $\ket{n}$ are the eigenstates of $\hat{H}_c $ and $\ket{n'}$ ($\ket{n''}$) are the first-order (second-order) corrections. 
Such terms modify significantly the structure of the wavepacket and are, therefore, included in the derivation of the equations of motion.
The wavepacket is, thus, constructed as:%from the perturbed eigenstates $\ket{\tilde{n}}$ as:
\begin{eqnarray}
	\ket{\tilde{W}_0} :=\displaystyle\int\limits_{\mathbb{T}^d} \mathrm{d}^d \mathrm{k} w(\bs{k},t)\sum\limits_{n}\eta_n  e^{i\tilde{\theta}^{nm}(\bs{k} , t)}  	\ket{\tilde{m} (\bs{k},\bs{r})}\,,
\end{eqnarray}    
where $ {w}(\bs{k},t) $ is now the distribution function, $|\eta_n |^2 = 1$ is the probability of a particle being in the $n$-th energy band, $\tilde{\theta}(\bs{k} , t) = \int_{0}^{t}\mathcal{{\tilde{E}}}^{nm}(\bs{k})\mathrm{d}t' + \tilde{\gamma}^{nm}(t)$ is the sum of the dynamical phase given by the temporal integral over the perturbed energy dispersion $ \mathcal{{\tilde{E}}}^{nm} = 	\bra{\tilde{n} (\bs{k},\bs{r})}{\hat{H}_0 + \hat{H}'}	\ket{\tilde{m} (\bs{k},\bs{r})}$, and the geometrical phase is $\tilde{\gamma}^{nm} = -i\int_{0}^{t} \tilde{\mathcal{A}}^{nm}_{t}\mathrm{d}t'$ with $\tilde{\mathcal{A}}^{nm}_{t} := i \bra{\tilde{n} (\bs{k})}\frac{\mathrm{d}}{\mathrm{d}t} \ket{\tilde{n} (\bs{k})}$.	
The center-of-mass position $\bs{r}$ and momentum $\bs{k}$ of this wavepacket is defined as~\cite{Sundaram1999}
\begin{eqnarray}
	 \bs{r} := \bra{W_0}\hat{\bs{r}}\ket{W_0} \equiv\Tr\underline{\bs r}\hspace{5pt}\&\hspace{5pt}\bs{k} :=  \bra{W_0}\hat{\bs{k}}\ket{W_0}\equiv\Tr\underline{\bs k} \,,\notag
	\label{eq:CoM conditions}
\end{eqnarray}  
where $\underline{\bm r}$ and $\underline{\bm k}$ are the matrix representations of the position and momentum operator.
In the case of a single occupied band, $\underline{\bm r}$ and $\underline{\bm k}$ become real numbers and the trace is omitted.

\section{Spinor-multipoles moments \label{apx:poles}}

% \subsection{A Wilson loop approach}
In this Section we review the definition of Wilson loops and nested Wilson loops, as well as defining their spinor analogues.
These are ultimately related to the spinor-multipole moments, e.g., the spinor-dipole, spinor-quadrupole and spinor-octapole moments.
% We show that T.R. symmetry forces such quantities to vanish for any value of $\bm \phi$.
% We define spin dipole and quadrupole moments and show that can take nonzero values,
% or even be quantised by spatial symmetries.
% We relate spin accumulation and transport to the modulation of these multipoles.

The Wilson loop, defined as
\begin{eqnarray}
	\mathcal{W}_\mu = e^{i \int \frac{d{\bm k}}{2\pi} \mathcal{A}_{\mu}}\,,
	\label{eq:WLoop}
\end{eqnarray}
is constructed by integrating the connection $\mathcal{A}^{nm} _{\mu} = \bra{\psi^n (\bm k)}{\partial_{k_\mu}}\ket{\psi^m (\bm k)}$ over the entire Brillouin zone.
Its eigenvalues are related to the electronic positions relative to the positively charged atomic centers, a.k.a. Wannier centers,
\begin{eqnarray}
	\mathcal{W}_\mu \ket{w^{n}_\mu (\bm k)} = e^{i 2\pi w^n_\mu }\ket{w^{n}_\mu (\bm k)}\,,
\end{eqnarray}
where $\{w^n_\mu\}$ is a set of Wannier centers in the $\mu$-direction with associated eigenvectors $ \ket{w^{n}_\mu (\bm k)} $.
The electric dipole moment density is determined by the displacement of electrons from their atomic centers, i.e.,
\begin{eqnarray}
	\mathcal{P}_\mu = 	-\frac{i}{2\pi } \log\det \mathcal{W}_\mu\,.
\end{eqnarray}

% In general, the bulk electric dipole $\bm{\mathcal{P}}$ of a T.R. invariant insulator can be decomposed to two contributions originating from the two T.R.-partners, cf. Fig.~\ref{fig:fig1dpolcur},
% \begin{eqnarray}
% 	\bm{\mathcal{P}} = \bm{P}^1 + \bm{P}^2\,.
% \end{eqnarray}
% Here, $\bm{P}^{i}$ refers to the electric dipole moment of the $i$-th T.R.-sector.
% Since the system has T.R symmetry, each contribution comes with an opposite sign, hence, their sum vanishes $\bm{P}^1  =  -\bm{P}^2$.

The nested Wilson loop is defined as
\begin{eqnarray}
	\mathcal{W}^{\pm}_{\mu\nu} = e^{i \int \frac{d{\bm k}}{2\pi} \mathcal{A}^{\pm}_{\mu\nu}}\,.
\end{eqnarray} 
The connection $[\mathcal{A}^{\pm} _{\mu\nu} ]^{nm}= \bra{u_\nu ^n (\bm k)}{\partial_{k_\mu}}\ket{u_\nu ^m (\bm k)}$ is defined over the, so called, Wilson bands $\ket{u_\mu ^n  (\bm k)}=\sum_{i}  \ket{w_\mu ^n  (\bm k)}^i \ket{\psi^i (\bm k)}$, where $\ket{w_\mu ^n  (\bm k)}^i$ is the $i$-th component of the $\mathcal{W}_\mu$ Wilson loop eigenvector $\ket{w_\mu ^n  (\bm k)}$.
Here, the $\pm$ superscript denotes the Wannier sector that is comprised by either positive or negative eigenvalues.% In other words, it is the polarization of Wannier bands.
The electric quadrupole moment density is measured by the product of the averaged eigenvalues of nested Wilson loops, summed over the Wannier sectors
\begin{eqnarray}
	\mathcal{Q}_{\mu\nu} = \frac{1}{(2\pi )^2 } \sum\limits_{\sigma}\log\det \mathcal{W}^{\sigma}_{\mu\nu}\log\det\mathcal{W}^{\sigma}_{\nu\mu}\,.
\end{eqnarray}
% As before, T.R. symmetry allows for the decomposition of $\mathcal{Q}_{\mu\nu}$ into two contributions originating from the two spin sectors,
% \begin{eqnarray}
% 	\mathcal{Q}_{\mu\nu} = Q^1_{\mu\nu} + Q^2_{\mu\nu}\,.
% \end{eqnarray}
% In the 2D model of \eqref{eq:2DHam PS}, each contribution $Q^i _{\mu\nu}$ takes nonzero values, however, their sum always vanishes due to T.R. symmetry that relates $Q^1_{\mu\nu} =- Q^2_{\mu\nu}$.

The nested-nested Wilson loop is defined as
\begin{eqnarray}
	\mathcal{W}^{\pm }_{\mu\nu\rho}= e^{i \int \frac{d{\bm k}}{2\pi} \mathcal{A}^\pm _{\mu\nu\rho}}\,,\\\notag
\end{eqnarray} 
The connection $[\mathcal{A}^{\pm} _{\mu\nu\rho} ]^{nm}= \bra{u_{\nu\rho} ^n (\bm k)}{\partial_{k_\mu}}\ket{u_{\nu\rho} ^m (\bm k)}$ is defined over the nested-Wilson bands $\ket{u_{{\nu\rho}} ^n  (\bm k)}=\sum_{i}  \ket{w_{\nu\rho} ^n  (\bm k)}^i \ket{u_\rho^i (\bm k)}$, where $\ket{w_{\nu\rho} ^n  (\bm k)}^i$ is the $i$-th component of the $\mathcal{W}_{\nu\rho}$ nested-Wilson loop eigenvector $\ket{w_{\nu\rho} ^n  (\bm k)}$.
The $\pm$ superscript denotes the nested-Wannier sector that is comprised by either positive or negative eigenvalues of both Wilson and nested-Wilson loops.
The octapole moment is calculated by
\begin{eqnarray}
	\mathcal{O}_{\mu\nu\rho} = \frac{i}{(2\pi)^3 } \sum\limits_{\sigma}\log \det\mathcal{W}^{\sigma }_{\mu\nu\rho}\log\det\mathcal{W}^{\sigma }_{\rho\mu\nu}\log\det \mathcal{W}^{\sigma}_{\nu\rho\mu}\,.\notag\\
\end{eqnarray}

We construct the spinor analogue of Wilson loop as
\begin{eqnarray}
	\mathcal{W}^{\mathcal{\hat{S}}} _\mu = e^{i \int \frac{d{\bm k}}{2\pi} S_\mu \mathcal{A}_{\mu}}\,,
\end{eqnarray}
where $S ^{nm} _\mu = \bra{u_\mu ^n  (\bm k)}\mathcal{\hat{S}}\ket{u_\mu ^m  (\bm k)}$ are the components of the spinor operator $\mathcal{\hat{S}}$ in the basis of Wilson bands.
% In fact, $S_\mu$ is diagonal due to a.T.R. symmetry. 
The spinor-dipole moment is given by
\begin{eqnarray}
	\mathcal{P}^{\mathcal{\hat{S}}} _\mu = 	-\frac{i}{2\pi } \log\det \mathcal{W}^{\mathcal{\hat{S}}}_\mu\,.
\end{eqnarray}
% When the system has T.R. symmetry, the bulk spin-dipole ${\bm {\mathcal{P}}}^{\text{spin}}$ is decomposed into two contributions originating from the two T.R.-partners
% \begin{eqnarray}
% 	\bm{\mathcal{P}}^{\text{spin}} = \bm{P}^1  - \bm{P}^2\,.
% \end{eqnarray}
% Here, $\bm{P}^i$ refers to the dipole moment of the $i$-th spin sector.
% In the 2D model of \eqref{eq:2DHam PS}, each contribution vanishes individually $\bm{P}^1  =  \bm{P}^2=0$ due to T.R. symmetry, hence, the spin-dipole moment is zero.
% This is reflected to the vanishing edge spin accumulation and to the trivial \Fst spin-Chern number.

The spinor-nested Wilson loop is defined as
\begin{eqnarray}
	\mathcal{W}^{\pm,\mathcal{\hat{S}}  }_{\mu\nu}= e^{i \int \frac{d{\bm k}}{2\pi}S_{\mu\nu} \mathcal{A}^\pm _{\mu\nu}}\,,
\end{eqnarray} 
where $S ^{nm} _{\mu\nu} = \bra{u_{\mu\nu} ^n  (\bm k)}\mathcal{\hat{S}}\ket{u_{\mu\nu} ^m  (\bm k)}$ are the components of the spinor operator $\hat{S}$ in the basis of nested-Wilson bands $\ket{u_{\mu\nu} ^n  (\bm k)}=\sum_{i}  \ket{w_{\mu\nu} ^n  (\bm k)}^i \ket{u_\nu ^i (\bm k)}$, where $\ket{w_{\mu\nu} ^n  (\bm k)}^i$ is the $i$-th component of the $\mathcal{W} _{\mu\nu}$ nested Wilson loop eigenvector $\ket{w_{\mu\nu} ^n  (\bm k)}$.
The spinor-quadrupole moment is then given by
\begin{eqnarray}
	\mathcal{Q}^{\mathcal{\hat{S}}  }_{\mu\nu} = \frac{1}{(2\pi )^2} \sum\limits_{\sigma}\log \det\mathcal{W}^{\sigma,\mathcal{\hat{S}} }_{\mu\nu}\log\det\mathcal{W}^{\sigma,\mathcal{\hat{S}} }_{\nu\mu}\,.
\end{eqnarray}

% For T.R.-ivariant systems the 

Lastly, the spinor-nested-nested Wilson loop is defined as
\begin{eqnarray}
	\mathcal{W}^{\pm,\mathcal{\hat{S}}  }_{\mu\nu\rho}= e^{i \int \frac{d{\bm k}}{2\pi}S_{\mu\nu\rho} \mathcal{A}^\pm _{\mu\nu\rho}}\,,
\end{eqnarray} 
where $S ^{nm} _{\mu\nu\rho} = \bra{u_{\mu\nu\rho} ^n  (\bm k)}\mathcal{\hat{S}}\ket{u_{\mu\nu\rho} ^m  (\bm k)}$ are the components of the spinor operator $\hat{S}$ in the basis of nested-nested Wilson bands $\ket{u_{\mu\nu\rho} ^n  (\bm k)}=\sum_{i}  \ket{w_{\mu\nu\rho} ^n  (\bm k)}^i \ket{u_{\nu\rho} ^i (\bm k)}$, where $\ket{w_{\mu\nu\rho} ^n  (\bm k)}^i$ is the $i$-th component of the $\mathcal{W} _{\mu\nu\rho}$ nested-nested Wilson loop eigenvector $\ket{w_{\mu\nu\rho} ^n  (\bm k)}$.
The spinor-octapole moment is calculated by
\begin{eqnarray}
	\mathcal{O}^{\mathcal{\hat{S}}  }_{\mu\nu\rho} = \frac{i}{(2\pi)^3 } \sum\limits_{\sigma}\log\det \mathcal{W}^{\sigma,\mathcal{\hat{S}} }_{\mu\nu\rho}\log\det\mathcal{W}^{\sigma,\mathcal{\hat{S}} }_{\rho\mu\nu}\log \det\mathcal{W}^{\sigma,\mathcal{\hat{S}} }_{\nu\rho\mu}\,.\notag\\
\end{eqnarray}

% \subsection{A geometrical approach}
% We use Eqs.~\eqref{} to give a geometrical definition of the spinor-multipoles in the case where the gauge freedom is fixed by boundary conditions.
% First, the spinot-dipole moment is given by
% \begin{eqnarray}
%     \mathcal{\bm P}^{\hS} _i = \mathcal{A}_{k_i} \left(1 +\epsilon_{i,j} \partial_{r_j} \left(\mathcal{A}_{k_j}\left(1+\epsilon_{j,l} \partial_{r_l} \mathcal{A}_{k_l}\right) +\epsilon_{j,l}\partial_{r_l} \mathcal{A}_{k_l}\right)\right)\notag\\
% \end{eqnarray}

% \bibliographystyle{ieeetr}
% \bibliography{ref.bib}

\end{document}